\documentclass[aps,prd,reprint,superscriptaddress,nofootinbib,eqsecnum,preprintnumbers]{revtex4-1}

\usepackage[T1]{fontenc}
\usepackage{newtxtext}
\usepackage[varg]{newtxmath}
\usepackage{esint}

\usepackage{titlesec}

\newcommand{\mysectionnumbering}{\thesection.~}

\titleformat{\section}{\bfseries\center\uppercase}{\mysectionnumbering}{0em}{}
\titleformat{\subsection}{\bfseries\center}{}{0.1em}{\thesubsection.~}
\titleformat{\subsubsection}{\bfseries\itshape\center}{}{0.1em}{\thesubsubsection.~}
\titleformat{\paragraph}[runin]{\itshape}{}{0em}{(\theparagraph)\hspace{0.5em}}[.]

\titlespacing{\section}{0pt}{2em plus 0.1em minus 0.1em}{0.7em}
\titlespacing{\subsection}{0pt}{1.5em}{0.5em}
\titlespacing{\subsubsection}{0pt}{1.5em}{0.5em}

\let\oldappendix\appendix

\makeatletter
\renewcommand{\appendix}{%
\@ifstar
\oneappendix
\manyappendices
}
\makeatother

\newcommand{\oneappendix}{%
\oldappendix*
\renewcommand{\mysectionnumbering}{\MakeUppercase{Appendix}:~}
\renewcommand\theequation{\Alph{section}\arabic{equation}}
}
\newcommand{\manyappendices}{%
\oldappendix
\renewcommand{\mysectionnumbering}{\MakeUppercase{Appendix}~\thesection:~}
\renewcommand\theequation{\Alph{section}\arabic{equation}}
}

\makeatletter
\renewcommand\@makefntext[1]{%
\noindent{\hspace{1em}}{\@makefnmark}#1}
\makeatother

\makeatletter
\renewcommand\@makefnmark{\hbox{\@textsuperscript{\normalfont\color{black}\@thefnmark}}}
\makeatother

\renewcommand{\footnoterule}{%
  \kern -3pt
  \hrule width 1.2cm
  \kern 4pt
}

\renewcommand{\dot}[1]{\overset{\bm .}{#1}}

\newcommand{\eqskip}{{\newpage~\vskip-2.5\baselineskip}}

\usepackage[dvipsnames]{xcolor}
\usepackage[colorlinks=true,allcolors=Blue,breaklinks=true,hyperfootnotes=false]{hyperref}

\usepackage{enumitem}
\setlist[enumerate]{leftmargin=2.7em,topsep={0pt},label={(\arabic*)},itemsep=0pt,parsep=0em}

\usepackage{graphicx}

\usepackage{bm}

\let\bmf\undefined
\renewcommand{\hbar}{\hslash}
\newcommand{\bmf}[1]{\mathbf{#1}}
\newcommand{\dx}{\textup{d}}
\newcommand{\w}{\omega}
\newcommand{\mpl}{\ensuremath{m_\text{Pl}}}
\newcommand{\avg}[1]{\langle #1 \rangle}
\DeclareMathOperator{\Real}{Re}
\DeclareMathOperator{\Imaginary}{Im}
\newcommand{\X}{\mathcal{X}}
\newcommand{\Q}{\mathcal{Q}}
\newcommand{\A}{A} 
\newcommand{\K}{\kappa} 
\newcommand{\eps}{\varepsilon} 
\usepackage{xspace}
\newcommand{\dof}{d.o.f.\xspace}

\DeclareSymbolFont{cmodern}{OML}{cmm}{m}{i}
\DeclareMathSymbol{\partial}{0}{cmodern}{64}
\DeclareMathSymbol{g}{\mathalpha}{cmodern}{103}
\DeclareMathAlphabet{\mathcal}{OMS}{cmsy}{m}{n}
\DeclareFontFamily{U}{futm}{}
\DeclareFontShape{U}{futm}{m}{n}{<-> fourier-bb}{}
\DeclareMathAlphabet{\mathbb}{U}{futm}{m}{n}

\begin{document}

\title{Effective field theory for black holes with induced scalar charges}

\author{Leong Khim \surname{Wong}}

\author{Anne-Christine \surname{Davis}}
\affiliation{DAMTP, Centre for Mathematical Sciences, University of Cambridge, Wilberforce Road, Cambridge CB3 0WA, United Kingdom}

\author{Ruth \surname{Gregory}}
\affiliation{Centre for Particle Theory, Durham University, South Road, Durham DH1 3LE, United Kingdom}
\affiliation{Perimeter Institute, 31 Caroline Street North, Waterloo, Ontario N2L 2Y5, Canada}

\date{July 9, 2019}

\preprint{DAMTP-2019-9, DCPT-19/05}

\begin{abstract}
While no-hair theorems forbid isolated black holes from possessing permanent moments beyond their mass, electric charge, and angular momentum, research over the past two decades has demonstrated that a black hole interacting with a time-dependent background scalar field will gain an induced scalar charge. In this paper, we study this phenomenon from an effective field theory (EFT) perspective. We employ a novel approach to constructing the effective point-particle action for the black hole by integrating out a set of composite operators localized on its worldline. This procedure, carried out using the in-in formalism, enables a systematic accounting of both conservative and dissipative effects associated with the black hole's horizon at the level of the action. We show that the induced scalar charge is inextricably linked to accretion of the background environment, as both effects stem from the same parent term in the effective action. The charge, in turn, implies that a black hole can radiate scalar waves and will also experience a ``fifth force.'' Our EFT correctly reproduces known results in the literature for massless scalars, but now also generalizes to massive real scalar fields, allowing us to consider a wider range of scenarios of astrophysical interest. As an example, we use our EFT to study the early inspiral of a black hole binary embedded in a fuzzy dark matter halo. 
\end{abstract}

\maketitle

\section{Introduction}

The uniqueness theorems pioneered by Israel \cite{Israel:1967wq} (see also Ref.~\cite{Chrusciel:2012jk} for a recent review) tell us that black holes are remarkably simple objects characterized only by their mass, electric charge, and angular momentum. Even if one considers more general field theories interacting with gravity, the general rule, summarized by the ``no-hair'' theorems 
\cite{Ruffini:1971bza,Bekenstein:1971hc,Hawking:1972qk,1971ApJ...166L..35T,Adler:1978dp,Zannias:1994jf,Bekenstein:1995un,Saa:1996aw,Sotiriou:2011dz,Chrusciel:2012jk,Graham:2014mda,*Graham:2014mda_E},
is that there are no additional charges that a black hole can carry.
These precise theorems are predicated on several crucial assumptions, however, which if violated can lead to a variety of new solutions. Many such examples are known today, including colored black holes \cite{Bizon:1990sr}, black holes with a cosmic string \cite{Achucarro:1995nu}, and black holes with complex massive scalar or Proca hair supported by rotation \cite{Herdeiro:2014goa,Herdeiro:2016tmi}, or real scalar hair supported by exotic gravitational couplings \cite{Yagi:2012ya,*Yagi:2012ya_E,
Sotiriou:2014pfa}.

In this paper, we revisit a different kind of circumvention of the no-hair theorems. By relaxing the assumptions of stationarity and asymptotic flatness, which bear little resemblance to astrophysical environments, even a minimally coupled, real scalar field can exhibit interesting phenomenology around a black hole. A classic example is the inflaton. Neglecting backreaction, Jacobson \cite{Jacobson:1999vr} showed that the solution near the event horizon\footnote{Solutions that extend all the way to the cosmological horizon have also been found \cite{Chadburn:2013mta,Gregory:2018ghc}, albeit only for spherical black holes.} is given by the Kerr metric surrounded by an effectively massless scalar,
\begin{equation}
\phi(t,r) = \phi_0 + \dot\phi_0
\left[ t + \frac{2GM r_+}{r_+ - r_-} \log \left(\frac{r-r_+}{r-r_-}\right)\right],
\end{equation}
where $\phi_0 + \dot\phi_0 t$ is the background ``coasting'' solution and $r_\pm$ mark the locations of the inner and outer horizons in Boyer-Lindquist coordinates. Although valid only while $\dot\phi_0 t$ is sufficiently small, this solution nonetheless remains a good effective description of the inflationary epoch from the perspective of the black hole, whose light-crossing time is much shorter than cosmological timescales.

Let us now zoom out on this solution by expanding in powers of $GM/r$. We can write
\begin{equation}
\label{eq:intro_Jacobson_phi_decomposition}
\phi = \Phi + \frac{Q}{4\pi r} + \mathcal O(1/r^2), \quad
Q = -\A\partial_t\Phi,
\end{equation}
where ${\A = 8\pi GM r_+}$ is the area of the event horizon. The first term, ${\Phi = \phi_0 + \dot\phi_0 t}$, describes the background scalar field that persists independently of the black hole. The effect of the black hole is to ``drag'' the scalar, leading to the Coulomb-like potential in the second term, whose dimensionless numerator $Q$ is called the \emph{induced scalar charge}.\footnote{Other definitions in the literature differ on minus signs and factors of $4\pi G$. We find this definition the most natural.}

More recently, Horbatsch and Burgess \cite{Horbatsch:2011ye} applied this result to models of the Universe in which late-time acceleration is driven by a rolling scalar. In such cases, all black holes should be dressed with a charge ${Q = - \A\dot\phi_0}$, which they argue enables a black hole to radiate energy and momentum into scalar waves. Furthermore, as scalar radiation would lead to a faster decay in the orbital period of a binary, they arrive at the constraint
\begin{equation}
\sqrt{4\pi G}|\dot\phi_0| \lesssim (16~\text{days})^{-1} 
= 7 \times 10^{-7}~\text{s}^{-1}
\end{equation}
on any rolling scalar in the vicinity of the quasar OJ287. This bound stems from the supermassive black hole binary at the center of the quasar having an inspiral consistent with the predictions of general relativity in vacuum to within an uncertainty of 6\% \cite{Sillanpaa:1988zz,Valtonen:2008tx,Dey:2018mjg}. While by no means a spectacular bound (a slow-rolling scalar should satisfy ${\sqrt{4\pi G}|\dot\phi_0| \ll H_0 = {2 \times 10^{-18}~\text{s}^{-1}}}$), that black holes are sensitive to this value at all is interesting. Black holes observed by LIGO have also been used to constrain this effect \cite{Yunes:2016jcc}, but the bound obtained is much looser.

It is worth emphasizing that this behavior is not unique to rolling scalars: Black holes will develop scalar charges when embedded in \emph{any} arbitrary scalar-field environment, as long as the background scalar evolves in time relative to the black hole's rest frame. This intuition is supported by numerical relativity simulations \cite{Healy:2011ef,Berti:2013gfa}, 
which show that scalar radiation is also emitted by black holes moving through background scalar fields (even static ones) that are spatially inhomogeneous. In general, an analytic description of such systems is not possible, except when there exists a large hierarchy between the length and timescales of the black hole and its environment. In this limit, which will be our focus, the black hole can be approximated as a point particle traveling along the worldline of some effective center-of-energy 
coordinate. The general definition for the scalar charge should then be
\begin{equation}
\label{eq:def_Q}
Q(\tau) \coloneq - \A\dot\Phi\bm(z(\tau)\bm),
\end{equation}
where $\tau$ is the proper time along the worldline $z^\mu(\tau)$.

This brings us to the motivation for this work: Can we understand the full extent to which generic scalar-field environments affect the motion of black holes embedded within them? To date, only the flux of scalar radiation has been studied, but it is possible that a black hole's scalar charge impacts the inspiral in other ways. Moreover, previous analytic studies have all been 
limited to massless scalar-field backgrounds varying at most linearly with space and time. In this case, results can be obtained by appropriating Damour and Esposito-Far\`{e}se's calculations \cite{Damour:1992we} for the inspiral of binary neutron stars in scalar-tensor theories, since the derivatives of $\Phi$ are constant. New ``technology'' will have to be developed, however, for backgrounds that are more complicated functions of space and time. This generalization is worth exploring, since many scenarios beyond the Standard Model predict the existence of massive (pseudo)scalar fields that can form localized, gravitationally bound objects, which resist collapse by oscillating in time \cite{Feinblum:1968nwc,Jetzer:1991jr,PhysRevLett.66.1659,Liebling:2012fv,Visinelli:2017ooc}. The prime example is a galactic fuzzy dark matter halo formed by an ultralight scalar of mass $\mu \sim 10^{-22}$--$10^{-21}~\text{eV}$ \cite{Hu:2000ke,Berezhiani:2015pia,Marsh:2015xka,Hui:2016ltb}. Whether black holes can be used to probe such configurations is an interesting question. (Of course, black hole superradiance already provides a way of probing new fundamental fields \cite{Brito:2015oca,Arvanitaki:2010sy,Arvanitaki:2014wva,Arvanitaki:2016qwi,Yoshino:2013ofa,*Yoshino:2014wwa,Brito:2014wla,Brito:2017wnc,*Brito:2017zvb,Hannuksela:2018izj,Cardoso:2018tly,Stott:2018opm,Baumann:2018vus}. Our work explores a complementary avenue, as it does not rely on rotation and only pertains to fields with a Compton wavelength much larger than the black hole.)

We push forward by constructing an effective field theory (EFT) \emph{\`{a} la} Goldberger and Rothstein \cite{Goldberger:2004jt,Goldberger:2005cd}, which describes black holes in terms of worldlines furnished with composite operators that capture finite-size effects. The key benefit of this description is the ability to disentangle questions about the long-distance, infrared (IR) physics we are interested in---such as the trajectory of the black hole---from the short-distance, ultraviolet (UV) physics transpiring near its horizon. Information about the latter is accessible to distant observers, like ourselves, through the way it impacts the black hole's multipolar structure. Mathematically, this is characterized in the EFT by Wilsonian coefficients, whose values can be determined by matching calculations with the ``full theory.'' As we are doing purely classical physics, we have the advantage of knowing what this UV completion is---it is just general relativity. 

This paper is organized as follows: We begin in Sec.~\ref{sec:kg} by solving perturbatively the Einstein-Klein-Gordon field equations for a black hole interacting with a massive scalar field. This generalizes Jacobson's result and will be later used to fix Wilsonian coefficients. We then construct the EFT in Sec.~\ref{sec:eft}. The main novelty of our approach is the way we obtain the black hole's effective action: By integrating out composite operators localized on its worldline using the in-in formalism, we obtain an action expressed in terms of correlation functions that can systematically account for both conservative and dissipative effects. Contained in these correlation functions are the aforementioned Wilsonian coefficients. We find that the coefficient responsible for the induced scalar charge also sets the accretion rate of the background scalar onto the black hole. This inextricable connection is the EFT's way of saying that the charge arises as a natural consequence of ingoing boundary conditions at the horizon.

The remainder of the paper is concerned with exploring our EFT's broader phenomenological implications. Section~\ref{sec:motion} presents the derivation of the universal part of the equation of motion for the black hole's worldline, demonstrating that the black hole experiences a drag force due to accretion and a fifth force due to its scalar charge. We then specialize to the case of a black hole binary embedded in a fuzzy dark matter halo in Sec.~\ref{sec:bin}. In addition to the effects already discussed in earlier sections, our EFT also provides a natural language for calculating two other effects not unique to black holes but common to any massive body: dynamical friction and the gravitational force exerted by the halo. Finally, our calculations are combined with observations of OJ287 to constrain the allowed local density of fuzzy dark matter. The result is a very weak upper bound, which is unsurprising, since typical halos are too dilute to leave any observable imprints in the binary's inspiral. The paper concludes in Sec.~\ref{sec:conclusion}, where we discuss some potential future applications of our EFT, which may lead to better observational prospects. Note that while we use the usual $\hbar=c=1$ units (except in Sec.~\ref{sec:kg} where we also set $G=1$), in this paper the reduced Planck mass is defined by $\mpl^2 = 1/(32\pi G)$ to be consistent with the EFT literature.

\section{Scalar multipole moments\protect\\in the full theory}
\label{sec:kg}

We start by considering what happens when a black hole of mass~$M$ is embedded within a background environment comprised solely of a Klein-Gordon field~$\phi$ of mass $\mu$. The problem is analytically tractable under four conditions:
\begin{enumerate}
\item As perceived by an observer in the rest frame of the black hole, the timescale $\w^{-1}$ on which the background varies is much longer than the black hole light-crossing time, $M\w \ll 1$.

\item Similarly, the background is assumed to vary on a length scale $\mathfrak R$ that is much greater than the black hole's radius, $M/\mathfrak R \ll 1$. 

\item The Compton wavelength $\mu^{-1}$ of the scalar is also assumed to be much greater than the size of the black hole, $M\mu \ll 1$.

\item The energy density in the scalar field is dilute enough that, in the immediate vicinity of the black hole, its backreaction onto the geometry is subdominant to the black hole's own spacetime curvature.
\end{enumerate}
Rather than being seen as just simplifying assumptions, these should be considered defining characteristics for what it means to be a background environment.

The last condition implies that the scalar behaves like a test field near the horizon of the black hole. By neglecting its backreaction, the problem of studying the effect of the black hole on $\phi$ reduces to one of solving the Klein-Gordon equation on a fixed Kerr background. This equation is separable in Boyer-Lindquist coordinates $(t,r,\theta,\varphi)$; thus, one can make the ansatz~\cite{Rowan:1977zg,Detweiler:1980uk}
\begin{equation*}
\phi \propto e^{-i \w t + i m \varphi} R_{\ell m}(r) S_{\ell m}(\theta),
\end{equation*}
where the integers $(\ell,m)$ label different angular-momentum states. To obtain an analytic solution, we further restrict attention to near-horizon distances\footnote{This suffices for our purposes, since larger distances are well within the purview of our EFT. We only need this full-theory calculation to resolve the UV physics near the horizon.} $r \ll \max 
(\w^{-1},\mu^{-1})$ and truncate the solution to first order in $M\w$ and $M\mu$. With these simplifications, the angular part of the solution $S_{\ell m}(\theta) e^{im\varphi}$ reduces to the spherical harmonics $Y_\ell^m(\theta,\varphi)$, while the radial part is \cite{Detweiler:1980uk}
\begin{gather}
R_{\ell m}(r) \propto \left(\frac{r - r_+}{r - r_-}\right)^{iP_m}
\hspace{-0.3em}
{}_2F{}_1\left( -\ell,\ell+1;1-2iP_m;\frac{r - r_-}{r_+ - r_-} \right),
\nonumber\\
\label{eq:kg_R_sol}
\end{gather}
having imposed ingoing boundary conditions at the future event horizon. The parameter $P_m$ is defined to be
\begin{equation}
P_m \coloneq \frac{am - 2M r_+ \w}{r_+ - r_-},
\end{equation}
where $a$ is the specific angular momentum of the black hole.

As we did with Jacobson's result, let us zoom out on Eq.~\eqref{eq:kg_R_sol} to obtain a coarse-grained description valid at distances ${M \ll r \ll \max(\w^{-1},\mu^{-1})}$. The two dominant terms are
\begin{equation}
\label{eq:kg_R_sol_large_r}
R_{\ell m}(r) \propto r^\ell + C_{\ell m}r^{-\ell-1},
\end{equation}
with relative coefficients, accurate to first order in $M\w$ and $M\mu$, given by
\cite{Detweiler:1980uk,Pani:2012bp}
\begin{equation}
\label{eq:kg_Clm}
C_{\ell m} = - iP_m (r_+ - r_-)^{2\ell+1} \frac{(\ell!)^2}{(2\ell)!(2\ell+1)!} 
\prod_{j=1}^\ell\big(j^2 + 4P_m^2\big).
\end{equation}
These expressions can now be used to read off a black hole's scalar multipole moments.

\subsection{Scalar charge}

Consider first the ${\ell = 0}$ mode. At distances $M \ll r \ll \max(\w^{-1},\mu^{-1})$, the solution reads
\begin{equation}
\label{eq:kg_sol_l=0}
\phi = \Phi_0 e^{-i\w t}( 1 + C_{00}r^{-1} ),
\end{equation}
having included an overall amplitude $\Phi_0$ for the field. From Eq.~\eqref{eq:kg_Clm}, $C_{00} = 2M r_+ i \w$, and for real scalar fields, taking the real part of Eq.~\eqref{eq:kg_sol_l=0} yields
\begin{equation}
\phi = \Phi_0 \cos \w t + \frac{8\pi M r_+ \Phi_0 \sin \w t}{4\pi r}.
\end{equation}
It should be readily apparent that this reproduces Eq.~\eqref{eq:intro_Jacobson_phi_decomposition}: For a background environment of the form $\Phi = \Phi_0 \cos \w t$ in the vicinity of the black hole, the full scalar field behaves as $\phi = \Phi + Q/(4\pi r)$, with the scalar charge $Q$ defined by Eq.~\eqref{eq:def_Q} as before.

\subsection{Higher multipole moments}

Now suppose our scalar field is not quite homogeneous but has a linear gradient: $\Phi = (\bmf b \cdot \bmf x) \cos \w t$. This induces a dipole moment in the scalar, via the $\ell = 1$ mode, whose solution is
\begin{equation}
\label{eq:kg_sol_l=1}
\phi = \sum_{m=-1}^1 b_m e^{-i\w t}(r + C_{1m} r^{-2}) Y_1^m(\theta,\varphi).
\end{equation}
The constants ${b_m \sim \mathcal O(1/\mathfrak R)}$ are related to the Cartesian components of the vector $\bmf b = (b_x,b_y,b_z)$ via
\begin{equation}
b_{\pm 1} = \sqrt{\frac{2\pi}{3}}(b_x + ib_y), \quad
b_0 = \sqrt{\frac{4\pi}{3}} b_z.
\end{equation}
Unlike the $\ell = 0$ case, $C_{1 m}$ has a term that is independent of~$\w$:
\begin{equation}
C_{1 m} = - \frac{i}{3} a m M^2 + \mathcal O(M\w).	
\end{equation}
Substituting this back into Eq.~\eqref{eq:kg_sol_l=1} reveals that in the presence of a nontrivial background scalar gradient $\nabla\Phi = \bmf b \cos\w t$, black holes also acquire a spin-dependent dipole moment,
\begin{equation}
\phi = \Phi + \frac{\bmf p \cdot \hat{\bmf x}}{4\pi r^2}, \quad
\bmf p = \frac{4\pi a M^2}{3}(\hat{\bmf S}\times\nabla\Phi) + \mathcal O(M\w),
\end{equation}
where~$\hat{\bmf S}$ is the unit vector along the black hole's spin axis. Notice that the dipole moment $\bmf p$ survives in the static limit ${\w \to 0}$. The no-hair theorems are still circumvented here because a linear spatial gradient ${\Phi \sim \bmf b\cdot\bmf x}$ violates the assumption of asymptotic flatness.

Spherical black holes can also attain higher-order moments, although the effect is suppressed by one power of $M\w$ relative to the spinning case. Setting $a=0$ in Eq.~\eqref{eq:kg_Clm} yields
\begin{equation}
\label{eq:kg_Clm_spherical}
\left.C_{\ell m}\right|_{a=0}
= \frac{(\ell!)^4}{(2\ell)!(2\ell+1)!} (2M)^{2\ell+2} i\w.
\end{equation}
Upon substitution into Eq.~\eqref{eq:kg_sol_l=1}, we find that the spin-independent part of the dipole moment is
\begin{equation}
\label{eq:def_spinless_dipole_moment}
\left.\bmf p \right|_{a=0} = - \frac{16\pi M^4}{3} \frac{\dx}{\dx t}(\nabla\Phi).
\end{equation}

The same procedure can be repeated for $\ell \geq 2$; hence, we learn that a black hole gains not just a scalar charge when immersed in an arbitrary scalar-field environment $\Phi(t,\bmf x)$, but an infinite set of multipole moments. In practice, however, it often suffices to keep only the scalar charge and, in the case of rotating black holes, the  spin-dependent dipole moment. Higher multipole moments are suppressed by ever greater powers of $M/\mathfrak R$, making their phenomenology increasingly irrelevant.


\section{The effective action}
\label{sec:eft}

The systems of interest in this paper are all governed by the action\footnote{Note that we write $\int_x = \int\dx^dx$ as shorthand. Later, we will also write $\int_p = \int\dx^dp/(2\pi)^d$ for integrals over momentum variables.}
\begin{equation}
\label{eq:eft_S_full}
S_f[g,\phi] = \int_x\sqrt{-g} \left(2 \mpl^2 R - \frac{1}{2}(\partial\phi)^2 - \frac{1}{2}\mu^2\phi^2\right).
\end{equation}
When the length and timescales of its environment are much greater than those of the black hole, the latter can be approximated as an effective point particle traveling along a worldline $z^\mu(\tau)$ with 4-velocity $u^\mu$, normalized to satisfy $u^\mu u_\mu = -1$. This description emerges after integrating out short-wavelength modes from the full theory to generate the effective action~\cite{Goldberger:2004jt}
\begin{equation}
\label{eq:eft_S}
S = S_f[g,\phi] + S_p[z,g,\phi].
\end{equation}
The first term $S_f$ now governs only the remaining long-wavelength modes of the fields $(g,\phi)$, while the dynamics of the worldline and its interaction with the fields living in the bulk are given by the point-particle action $S_p$.

Performing this integration generally leads to an infinite number of terms in $S_p$, which can be organized according to relevancy as an expansion in three small ``separation-of-scale parameters,''
\begin{equation}
\label{eq:eft_small_parameters}
GM/\mathfrak R \ll 1, \quad
GM\w \ll 1, \quad
\text{and}\quad
GM\mu \ll 1.
\end{equation}
In this section, we discuss how to systematically construct $S_p$ and determine the most relevant terms needed to describe the interaction of a black hole with its scalar-field environment.

\subsection{Worldline degrees of freedom}
\label{sec:eft_dof}

Finite-size effects are modeled in the EFT by introducing a set of composite operators $\{ q^L(\tau), \,\dots \}$ localized on the worldline, which represent short-wavelength degrees of freedom (\dof)~living near the horizon \cite{Goldberger:2005cd,Kol:2008hc,*Kol:2008hc_v2,Endlich:2016jgc}. Using standard EFT reasoning, we then construct the effective action by writing down all possible terms that couple these operators to the long-wavelength fields $(g,\phi)$ in a way that is consistent with the symmetries of the theory. In this case, they  are general covariance, worldline reparametrization invariance, and worldline SO(3) invariance. (We restrict attention to spherical black holes for simplicity; the generalization to rotating ones is left for the future.) These steps lead us to the ``intermediary'' point-particle action
\eqskip
\begin{equation}
\label{eq:eft_Ip}
I_p = - \int_\tau M + \sum_{\ell=0}^\infty \int_\tau q^L(\tau)
\nabla_L \phi + \cdots.
\end{equation}
The first term is the familiar action for a point mass~$M$. The second term accounts for all possible interactions between the black hole and the real scalar field $\phi$. Analogous terms that couple other worldline operators to the curvature tensors are also present, but these have been omitted from Eq.~\eqref{eq:eft_Ip} and will be neglected in this paper, since they become important only at much higher orders in perturbation theory \cite{Goldberger:2004jt,Galley:2008ih}. Note that conventional multi-index notation is being used \cite{Blanchet:2013haa}:  The worldline operators are written as $q^L \equiv q^{\hat\imath_1 \dots \hat\imath_\ell}$, whereas $\nabla_L \equiv \nabla_{\hat\imath_1} \dots \nabla_{\hat\imath_\ell}$ denotes the action of multiple covariant derivatives. The indices $\hat\imath\in\{1,2,3\}$ label the three directions in the black hole's rest frame that are mutually orthonormal to one another and to the tangent $u^\mu$ of the worldline.

Traces of $\nabla_L\phi$ are redundant operators; hence, they can be absorbed into redefinitions of $q^{L-2n}$, where $n$ counts the number of traces \cite{Kol:2008hc,Endlich:2016jgc}. As a result, the worldline operators $q^L(\tau)$ can be taken to be symmetric and trace free (STF). The set of all STF tensors of rank $\ell$ generates an irreducible representation of SO(3) of weight~$\ell$ \cite{Thorne:1980ru}; thus, the worldline operators admit an interpretation as dynamical multipole moments of the black hole \cite{Goldberger:2005cd}. The $\ell=0$ operator $q(\tau)$ must therefore be responsible for the induced scalar charge, while the $\ell=1$ operator $q^{\hat\imath}(\tau)$ will lead to the induced dipole moment. The $\ell$th operator, in turn, corresponds to the $\ell$th multipole moment.

As its name suggests, the intermediary point-particle action \eqref{eq:eft_Ip} is not yet the end of the story. At the moment, it is comprised of both UV \dof, which a distant observer cannot directly probe, and the IR \dof $(z,g,\phi)$ that we ultimately care about. While it is possible to perform calculations directly with this action (see, e.g., Refs.~\cite{Goldberger:2005cd,Endlich:2016jgc}), for our purposes it will be instructive---and more convenient---to integrate out $q^L$ and obtain a truly effective point-particle action:
\begin{equation}
\label{eq:eft_Sp_integral}
S_p[z,g,\phi] = -i\log\int Dq^L \exp(i I_p[z,g,\phi,q^L]).	
\end{equation}
Being dynamical variables in their own right, the worldline operators $q^L$ come with kinetic terms that govern their dynamics, but we have also neglected to write these down explicitly in Eq.~\eqref{eq:eft_Ip} since their exact forms are unknown to us. Without detailed knowledge of their kinetic terms, integrating out $q^L$ leaves us with an effective action expressed in terms of their correlation functions $\avg{q^{L}(\tau) \dots q^{L'}(\tau')}$,\footnote{Expectation values are taken with respect to the ground state of the worldline theory, which corresponds to a classical, unperturbed black hole. Hawking radiation can be neglected.} which can be reconstructed through a series of matching calculations with the full theory. The situation simplifies tremendously, however, if we assume that the dynamics of these operators is fully characterized by their two-point correlation functions. Far from being just convenient, this assumption is linked to the test-field approximation in Sec.~\ref{sec:kg} and is thus valid under the conditions outlined therein.

\subsection{Integrating out}
\label{sec:eft_integration}

Because we are interested in studying the real, causal evolution of a system, rather than calculating in-out scattering amplitudes, the appropriate language required for integrating out the worldline operators is the in-in, or closed time path (CTP), formalism. (See Refs.~\cite{Schwinger1961,Keldysh1964,*Keldysh1964_v2,CHOU19851,PhysRevD.33.444,PhysRevD.35.495} for classic texts on the subject and Refs.~\cite{Galley:2008ih,Galley:2009px,Galley:2010xn,Birnholtz:2013nta} for applications similar to the present context.) At its heart, this formalism converts the standard version of Hamilton's variational principle, which is inherently a boundary value problem, into an initial value problem. It accomplishes this by doubling all dynamical \dof~$\Psi \to (\Psi_1,\Psi_2)$ and allowing the two copies to evolve independently subject to appropriate boundary conditions. Physical observables are obtained by making the identification $\Psi_1 = \Psi_2 = \Psi$ at the end. Following Galley~\cite{Galley:2012hx}, we will refer to this identification as ``taking the physical limit.''

\subsubsection{Fixed worldlines}
\label{sec:eft_integration_fixed_worldlines}

The \dof~of our EFT are $\Psi = \{ z^\mu, g_{\mu\nu},\phi, q^L \}$, and we wish to integrate out $q^L$. It will be instructive to begin by considering a simplified problem in which we fix the metric and worldline to be nondynamical. Under this restriction, the intermediary point-particle action \eqref{eq:eft_Ip} reads
\begin{equation}
\label{eq:eft_dissipative_Sp}
I_p = \int_\tau (q_1 \phi_1 - q_2 \phi_2) + \cdots
\end{equation}
when recast in the in-in formalism. We focus on the $\ell=0$ operator to streamline the discussion, although the generalization to higher multipole moments is straightforward. Introducing CTP indices $a,b \in\{1,2\}$ allows us to write
\begin{equation*}
q_1\phi_1 - q_2\phi_2 = c^{ab} q_a \phi_b = q_a\phi^a.
\end{equation*}
Note that all our \dof~$\Psi_a = (\Psi_1,\Psi_2)$ innately come with a downstairs index; indices are raised with the CTP metric $c^{ab} = c_{ab} = \text{diag}(1,-1)$.

The assumption that the dynamics of $q(\tau)$ is fully characterized by its two-point functions implies that Eq.~\eqref{eq:eft_Sp_integral} is a Gaussian integral that can be evaluated exactly to yield
\begin{equation}
S_p = \int_\tau \avg{q_a} \phi^a + \frac{1}{2} \int_{\tau,\tau'} 
\chi_{ab}(\tau,\tau')\phi^a(\tau) \phi^b(\tau').
\end{equation}
If nonvanishing, the vacuum expectation value $\avg{q_a}$ in the first term describes a permanent scalar charge of the black hole. From what we know of the no-hair theorems, this must be zero, leaving us with only the linear response in the second term. The matrix of two-point functions is~\cite{CHOU19851,PhysRevD.35.495}
\begin{equation}
\chi_{ab} = 
\begin{pmatrix}
\chi_F & \chi_- \\
\chi_+ & \chi_D
\end{pmatrix}	
\end{equation}
(see Appendix~\ref{sec:rules_worldline} for details on the individual two-point functions) and satisfies the symmetry property
\begin{equation}
\label{eq:def_chi_identity}
\chi_{ab}(\tau,\tau') = \chi_{ba}(\tau',\tau).
\end{equation}

In most circumstances, it is more convenient to work in a different basis called the Keldysh representation. Define the average and difference of our two copies as, respectively,
\begin{equation}
\label{eq:def_Keldysh_repn}
\Psi_+ \coloneq \frac{1}{2}(\Psi_1 + \Psi_2), \quad
\Psi_- \coloneq \Psi_1 - \Psi_2.
\end{equation}
In the physical limit (PL), $\Psi_+|_\text{PL}=\Psi$ and $\Psi_-|_\text{PL} = 0$. This transformation can also be written in index notation as
\begin{equation}
\label{eq:eft_Keldysh_transformation_rules}
\Psi_A = \Lambda_A{}^a\Psi_a, \quad
\Lambda_A{}^a = 
\begin{pmatrix}
\frac{1}{2} & \frac{1}{2} \\
1 & -1	
\end{pmatrix},
\end{equation}
with $A,B\in\{+,-\}$. Similarly, CTP tensors like $\chi_{ab}$ transform as $\chi_{AB} = \Lambda_A{}^a\Lambda_B{}^b\chi_{ab}$. Using the identities in Eq.~\eqref{eq:eft_chi_identities},
\begin{equation}
\label{eq:def_chi_AB}
\chi_{AB} =
\begin{pmatrix}
\frac{1}{2} \chi_H & \chi_R \\
\chi_A & 0
\end{pmatrix}.
\end{equation}
Because the transformation is linear, the identity in Eq.~\eqref{eq:def_chi_identity} holds also in this basis. Indices can still be raised and lowered with the CTP metric, which in this representation reads
\begin{equation}
c^{AB} = c_{AB} =
\begin{pmatrix}
0 & 1 \\
1 & 0
\end{pmatrix}.
\end{equation}

Repeating similar steps for the higher multipole moments and using the no-hair theorems to infer that $\avg{q^L} = 0$, in general we have
\begin{equation}
S_p = \frac{1}{2} \sum_{\ell=0}^\infty \int_{\tau,\tau'} \chi_{AB}^{LL'}(\tau,\tau')
\nabla_L\phi^A(\tau) \nabla_{L'}\phi^B(\tau').
\label{eq:eft_fw_Sp}
\end{equation}

\subsubsection{Dynamical worldlines}
\label{sec:eft_integration_dynamical_worldlines}

Having gained a sense for how this calculation proceeds, let us now integrate out $q^L$ in the general case when all our \dof~$\Psi = \{z^\mu,g_{\mu\nu},\phi,q^L\}$ are dynamical. Complications arise when there are two copies $(z_1,z_2)$ of the worldline for \emph{one} black hole, each with their own proper times, since the operators $q_1^L(\tau_1)$ appear to be living on the first copy $z_1(\tau_1)$, whereas $q_2^L(\tau_2)$ live on the second. How, then, should we integrate out these worldline operators, given that they appear to be living on different spaces?

The resolution comes by recalling that~$z^\mu$ are merely parametrizations in a given coordinate chart. The worldline itself is a map ${\gamma : \mathcal I \to \mathcal M}$ from the interval~$\mathcal I \subset \mathbb R$ to the bulk, four-dimensional manifold~$\mathcal M$. When there are two copies~$z_a$, there are also two maps~$\gamma_a$, but there is still only one underlying manifold~$\mathcal I$. Let $\lambda$ or $\sigma$ be the coordinate on $\mathcal I$ used to parametrize both copies of the worldline simultaneously. The tangent to each worldline is written as $\dot z{}^\mu_a = \dx z^\mu_a/\dx\lambda$. (We reserve $u^\mu$ for when the worldline is parametrized by its proper time.) The operators $q_a^L\equiv q_a^L(\tau_a)$ are pulled back onto $\mathcal I$ via the map
\begin{equation}
\tau_a(\lambda) = \tau_a(\lambda_i) + \int^\lambda_{\lambda_i} \dx\sigma
\sqrt{- g_{a,\mu\nu}\bm(z_a(\sigma)\bm) \frac{\dx z_a^\mu}{\dx\sigma} 
\frac{\dx z_a^\nu}{\dx\sigma}},
\end{equation}
where it should be understood that the CTP index $a$ above is acting as a placeholder and is not to be summed over. We are always free to choose the lower integration limit $\lambda_i$ and the initial value~$\tau_a(\lambda_i)$. The intermediary point-particle action thus reads
\begin{gather}
I_p = \int_\lambda \left[
-M \dot\tau_1(\lambda) + 
\dot\tau_1(\lambda)q_1\bm(\tau_1(\lambda)\bm) \phi_1\bm( z_1(\lambda)\bm)
\right]
- (1\leftrightarrow 2).
\label{eq:eft_dw_Sp_raw}
\end{gather}
As before, we focus only on the $\ell=0$ operator, since it is straightforward to generalize the following steps for $\ell \geq 1$.

Clearly, Eq.~\eqref{eq:eft_dw_Sp_raw} suggests we need better notation. To that end, we begin by generalizing the CTP metric to a set of tensors defined by
\begin{equation}
c^{a_1 \dots a_n} = 
\begin{cases}
+1 & a_1 = a_2 = \cdots = a_n =1, \\
-1 & a_1 = a_2 = \cdots = a_n =2, \\
\;0 & \text{otherwise}.
\end{cases}
\end{equation}
With these at our disposal, one can verify by direct evaluation that Eq.~\eqref{eq:eft_dw_Sp_raw} is equivalent to
\begin{equation}
I_p = -M\int_\lambda c^a\dot\tau_a(\lambda) 
+ \int_\lambda q_a(\lambda) \mathcal J^a(\lambda),
\label{eq:eft_dw_Sp_J}
\end{equation}
given sources $\mathcal J^a$ defined by
\begin{align}
\Delta^a(\lambda;x) &\coloneq
\int_\sigma c^{abcd}
\delta\bm(\lambda-\tau_b(\sigma)\bm) \,
\delta^{(4)}\bm(x - z_c(\sigma)\bm)\,
\dot\tau_d(\sigma),
\nonumber\\
\mathcal J^a(\lambda) &\coloneq
\int_x c^{abc} \Delta_b(\lambda;x) \phi_c(x).
\label{eq:eft_dw_def_J_Delta}
\end{align}
In this form, Eq.~\eqref{eq:eft_dw_Sp_J} is reminiscent of the simplified problem in Sec.~\ref{sec:eft_integration_fixed_worldlines}, apart from two minor differences: The manifold~$\mathcal I$ is parametrized by $\lambda$ rather than $\tau$, and the scalar field $\phi^a$ is here replaced by~$\mathcal J^a$. These prove to be no obstacle to 
evaluating the functional integral, which yields
\begin{gather}
S_p = -M \int_\lambda c^a\dot\tau_a(\lambda) + \frac{1}{2} \int_{\lambda,\lambda'}
\chi_{aa'}(\lambda,\lambda')\mathcal J^a(\lambda)\mathcal J^{a'}(\lambda').
\label{eq:eft_Sp_complete_CTP}
\end{gather}
Before proceeding any further, let us remark that the Hadamard propagator $\chi_H \equiv \chi_{++}$ appears in this action flanked by two powers of $\mathcal J^+ \equiv \mathcal J_-$, which vanishes in the physical limit. This implies that when we extremize the action $S$ to obtain the equations of motion for the system, $\chi_H$ will never contribute; thus, we set $\chi_H = 0$ from now on.

The hard work is over at this point, but the result in Eq.~\eqref{eq:eft_Sp_complete_CTP} is not yet written in a form convenient for calculations. Specifically, we want to make manifest its dependence on $\phi_a$. Using the definitions in Eq.~\eqref{eq:eft_dw_def_J_Delta}, we write
\begin{equation}
S_p = -M\int_\lambda c^a \dot\tau_a(\lambda) 
+ \frac{1}{2} \int_{x,x'} \X^{aa'}(x,x')\phi_a(x)\phi_{a'}(x'),
\label{eq:eft_integration_dynamical_worldlines_Sp_final}
\end{equation}
expressed in terms of the correlation functions
\begin{equation}
\X^{aa'}(x,x') \coloneq \int_{\lambda,\lambda'} c^{abc}c^{a'b'c'} 
\Delta_b\Delta_{b'}\chi_{cc'}.
\end{equation}
This is the desired end result. In the definition above, the two-point functions all depend on the same argument, $\chi_{cc'}\equiv\chi_{cc'}(\lambda,\lambda')$, and a primed index denotes dependence on primed variables, i.e., $\Delta_b\equiv\Delta_b(\lambda;x)$ whereas $\Delta_{b'}\equiv\Delta_{b'}(\lambda';x')$. As a generalization of Eq.~\eqref{eq:def_chi_identity}, it is easy to show that
\begin{equation}
\label{eq:def_X_identity}
\X^{aa'}(x,x') = \X^{a'a}(x',x).
\end{equation}

These correlation functions can be written in the Keldysh representation by utilizing the transformation rule
\begin{equation}
\X^{AA'}(x,x') = \X^{aa'}(x,x') \Lambda_a{}^A \Lambda_{a'}{}^{A'},
\end{equation}
where $\Lambda_a{}^A$ is the inverse of $\Lambda_A{}^a$, satisfying $\Lambda_a{}^{A} \Lambda_A{}^b = \delta_a^b$. Having explicit 
expressions for $\X^{AA'}$ will be useful. The same argument that allowed us to neglect $\chi_H$ earlier also allows us to neglect $\X^{--}$. Taken together with the symmetry property in Eq.~\eqref{eq:def_X_identity}, we conclude that it suffices to know only the following two components:
\begin{subequations}
\label{eq:eft_X_explicit}
\begin{align}
\X^{++}(x,x') =&\, \frac{1}{2}\int_{\lambda,\lambda'}
[2\chi_R(\Delta_1\Delta_{1'}-\Delta_2\Delta_{2'})
\nonumber\\
&- \chi_C(\Delta_1+\Delta_2)(\Delta_{1'}-\Delta_{2'})],
\label{eq:eft_X_explicit_++}
\allowdisplaybreaks\\
\nonumber\\
\X^{-+}(x,x') =&\, \frac{1}{4}\int_{\lambda,\lambda'}
[2\chi_R(\Delta_1\Delta_{1'}+\Delta_2\Delta_{2'})
\nonumber\\
&- \chi_C(\Delta_1-\Delta_2)(\Delta_{1'}-\Delta_{2'})].
\end{align}
\end{subequations}

In both cases, judicious use of Eqs.~\eqref{eq:eft_chi_definitions} and \eqref{eq:eft_chi_identities} has been made to express $\X^{AA'}$ only in terms of the retarded propagator $\chi_R$ and the commutator $\chi_C$ (and~$\chi_H$, which is then discarded). The definition in Eq.~\eqref{eq:def_chi_R} can then be used to infer that
\begin{equation}
\chi_R(\tau,\tau') - \chi_R(\tau',\tau) = \chi_C(\tau,\tau'),
\end{equation}
which in Fourier space reads\footnote{Since the worldline is reparametrization invariant, the two-point functions depend on $\tau$ and $\tau'$ only via their difference $\tau-\tau'$, making them amenable to a Fourier transform.}
\begin{equation}
\label{eq:fluctuation-dissipation-theorem}
\tilde\chi_C(\w) = \tilde\chi_R(\w) - \tilde\chi_R(-\w) = 2i\Imaginary\tilde\chi_R(\w). 
\end{equation}
Thus, once we know $\chi_R$, we also know $\chi_C$.

\subsection{Matching calculations}
\label{sec:eft_matching}

So far, the effective action we have constructed is fully generic and can account for finite-size effects of any spherical compact object interacting with a real scalar field. We will now specialize to black holes exclusively by fixing the form of the retarded propagator.

On general grounds, we expect $\chi_R$ to depend on both the black hole mass~$M$ and the scalar field mass~$\mu$. Dimensional analysis and the assumption of spherical symmetry are sufficient to deduce that
\begin{equation}
\tilde \chi^{LL'}_R(\w) = \delta^{LL'} (GM)^{2\ell+1} F_\ell(GM\w,GM\mu),
\end{equation}
where $\delta^{LL'}$ is the identity on the space of STF tensors of rank~$\ell$. The dimensionless functions $F_\ell$ almost certainly depend in a complicated way on their arguments. However, for low-frequency sources, we can expand in powers of the first argument to obtain
\begin{align}
F_\ell
&=
\sum_{n=0}^\infty \left[ F_\ell^{(2n)} (GM\w)^{2n} 
+ i F_\ell^{(2n+1)}(GM\w)^{2n+1} \right]
\nonumber\\
&=
F_\ell^{(0)} + i F_\ell^{(1)} GM\w + F_\ell^{(2)} (GM\w)^2 + \cdots,
\label{eq:eft_Fl_expansion}
\end{align}
where the dimensionless coefficients $F_\ell^{(n)}$ themselves admit an expansion in the remaining argument $GM\mu$. Naturally, the finite size of the black hole sets the UV cutoff for this EFT, and only the first few terms in this expansion are needed in practice when $GM\w\ll 1$. It is also worth remarking that this series cannot capture nonperturbative effects like quasinormal-mode resonances, but we do not expect such effects to be important in the low-frequency limit. The terms in Eq.~\eqref{eq:eft_Fl_expansion} even in $\w$ are time-reversal symmetric, and constitute what is called the ``reactive'' part of the black hole's response. On the other hand, the odd terms break time-reversal symmetry and are responsible for dissipative processes.

We can now determine the values of each of these Wilsonian coefficients by a matching calculation. To make contact with our results in Sec.~\ref{sec:kg}, we ought to compute field expectation values. While working with the full fields $(g,\phi)$ earlier was advantageous to manifestly preserve general covariance, to compute observables, we split
\begin{equation}
\label{eq:eft_matching_field_decomposition}
\phi_A = \Phi_A + \varphi_A, \quad
g_{A,\mu\nu} = \bar g_{A,\mu\nu} + \frac{h_{A,\mu\nu}}{\mpl}.
\end{equation}
The background fields $(\bar g_A,\Phi_A)$ describe a scalar-field environment that persists independently of the black hole. As these fields are nondynamical, we can immediately fix
\begin{equation*}
(\bar g_+,\Phi_+) = (\bar g,\Phi), \quad
(\bar g_-,\Phi_-) = 0.
\end{equation*}
Moreover, we will no longer have need to refer to the full metric explicitly, so let us drop the overbars and denote the background metric by~$g_{\mu\nu}$. Being much smaller than its environment, a black hole sources fluctuations $(h,\varphi)$ in the fields that can be treated perturbatively.

Expectation values of these fields can be computed by taking appropriate derivatives of the generating functional
\begin{align}
\mathcal Z[j,J] =& \int D h_\pm D\varphi_\pm \exp(iS[z,g+h/\mpl,\Phi+\varphi])
\nonumber\\
&\times \exp\left( i \int_x \sqrt{-g}(\varphi^A j_A + h^A_{\mu\nu} J_A^{\mu\nu} )\right),
\end{align}
where $(j_A, J_A^{\mu\nu})$ are arbitrary sources. This is approximated in perturbation theory by working with
\begin{equation}
\label{eq:eft_matching_Z[j,J]}
\mathcal Z[j,J] = \exp (i S_f^\text{(int)} + i S_p) \mathcal Z_0[j,J],
\end{equation}
where $\mathcal Z_0[j,J]$ is the (gauge-fixed) generating functional for the propagators of the free fields, and $S_f^\text{(int)}$ denotes the part of the field action not included in $\mathcal Z_0$. Further details can be found in Appendix~\ref{sec:rules_bulk}.

At leading order, $\avg{\varphi(x)}$ is sourced only by terms in $S_p$ that are linear in $\varphi$. Moreover, the worldline can be held fixed when computing field expectation values; hence, it suffices to work with the simplified action in Sec.~\ref{sec:eft_integration_fixed_worldlines}. Substituting the field decomposition \eqref{eq:eft_matching_field_decomposition} into the action~\eqref{eq:eft_fw_Sp}, we obtain
\begin{gather}
S_p = \sum_{\ell=0}^\infty \int_{\tau,\tau'} \chi_R^{LL'}(\tau,\tau')
\nabla_L \varphi_-(\tau)\nabla_{L'}\Phi(\tau') + \mathcal O(\varphi^2),
\end{gather}
having used Eq.~\eqref{eq:def_chi_identity} to simplify terms. Using the Fourier representation of $\chi_R$ and concentrating on the $F_0^{(1)}$ term for now, we find
\begin{align}
S_p &\supset
\int_{\tau,\tau'}\int_\w \left[  F_0^{(1)} (GM)^2 i\w + \cdots \right]
e^{-i \w (\tau-\tau')}\varphi_-(\tau) \Phi(\tau')
\nonumber\\
&=-\int_\tau F_0^{(1)} (GM)^2 \varphi_-(\tau) \dot\Phi(\tau) + \cdots,
\end{align}
where the second line follows from integrating by parts.

The Wilsonian coefficient $F_0^{(1)}$ characterizes the leading-order, low-frequency dissipative response and is responsible for the induced scalar charge of the black hole. To see this, we compute
\begin{align}
\avg{\varphi(x)} &= \avg{\varphi_+(x)}|_\text{PL}
\nonumber\\
&= \left.
(-i)^3\int_\tau (GM)^2 F_0^{(1)} \dot\Phi(\tau)
\frac{\delta^2 \mathcal Z_0[j,J]}{\delta j^+(x) \delta j^-\bm( z(\tau) \bm)}
\right|_{(j,J)=0}
\nonumber\\
&=
\frac{F_0^{(1)}}{16\pi} \int_\tau G_R\bm( x, z(\tau) \bm) Q(\tau),
\label{eq:eft_matching_Wilsonian_F01_GR}
\end{align}
where $G_R$ is the retarded propagator for the scalar field. The way this is written suggests that
\begin{equation}
F_0^{(1)} = 16\pi,
\label{eq:eft_matching_F01}
\end{equation}
and indeed this is true. We verify this by considering (so as to reproduce the scenario in Sec.~\ref{sec:kg}) a black hole at rest at the origin, $z^\mu(\tau) = (\tau,\bmf 0)$, around which the background field behaves as
\begin{equation*}
\Phi(x) = \Phi_0 \cos\w t + ({\bmf b}\cdot\bmf x)\cos \w t 
+ \mathcal O(r^2/\mathfrak R^2).
\end{equation*}
Recall that $\mathfrak R$ denotes a typical length scale of the background, $|{\bmf b}|\sim\mathcal O(1/\mathfrak R)$, and we will further assume a gravitationally bound state such that $\w^2 < \mu^2$. Moreover, let us suppose that $\Phi \sim\mathcal O(\eps)$ is sufficiently weak not just in the vicinity of the black hole but everywhere in spacetime, such that the background admits the weak-field expansion $g = \eta + H$, where $H \sim \mathcal O(\eps^2)$ is the backreaction of $\Phi$ onto the geometry. To leading order in $\eps$, it suffices to evaluate the integral in Eq.~\eqref{eq:eft_matching_Wilsonian_F01_GR} on flat space. Integrals of this form will need to be evaluated many times in this paper, and the general technique is reviewed in Appendix~\ref{sec:master_integral}. The result is 
\begin{equation}
\avg{\varphi(x)} = \frac{Q(t)}{4\pi r} e^{-\sqrt{\mu^2-\w^2} r},
\end{equation}
in total agreement with the full theory. Note that the Yukawa suppression is to be expected here, despite it not featuring in our results in Sec.~\ref{sec:kg}, since the latter concentrated only on distances $r \ll \max(\w^{-1},\mu^{-1})$. The same procedure can be repeated for the higher multipole moments; the spin-independent dipole moment in Eq.~\eqref{eq:def_spinless_dipole_moment}, for instance, is reproduced by our EFT provided
\begin{equation}
F_1^{(1)} = 16\pi/3.
\label{eq:eft_matching_F11}
\end{equation}

What about the other Wilsonian coefficients? To start with, consider the following three terms also present in the effective action:
\begin{align}
S_p \supset \int_\tau
& \big(
F_0^{(0)} GM \Phi - F_0^{(2)} (GM)^3 \ddot\Phi
\nonumber\\
&+
F_1^{(0)} (GM)^3 \partial^{\hat\imath}\Phi \partial_{\hat\imath}
+ \cdots\big)\varphi_-.
\label{eq:eft_matching_Sp_reactive}
\end{align}
These constitute the most relevant terms characterizing the reactive part of the response. Two comments are worth making at this stage: First, this part of the action could just as easily have been constructed by writing down all allowed contractions between $u^\mu$, the fields $(g,\phi)$, and their derivatives (see, e.g., Ref.~\cite{Damour:1998jk}). This bottom-up approach cannot account for dissipative processes, however, hence our more comprehensive and systematic route of integrating out worldline operators. Our second comment is that Eq.~\eqref{eq:eft_matching_Sp_reactive} is exactly the action Horbatsch and Burgess~\cite{Horbatsch:2011ye} took to be responsible for the induced scalar charge, but from what we have learned this cannot be true. The conclusions of their paper are nonetheless still valid, since their arguments do not rely on a specific form for the action.

Computing $\avg{\varphi(x)}$ as before, we find that the $F_0^{(0)}$ term generates a scalar-field profile due to a charge proportional to~$\Phi$, whereas the $F_1^{(0)}$ coefficient is responsible for a dipole moment proportional to $\partial_i\Phi$. Neither of these features are present in the full theory; thus, demanding consistency with the predictions of general relativity forces us to conclude that $F_0^{(0)} = F_1^{(0)}= 0$. More precisely, these coefficients are zero up to possible quadratic-order corrections in $GM\mu$, since our calculations in Sec.~\ref{sec:kg} are accurate only to linear order in $GM\mu$ and $GM\w$. Accordingly, the value of the coefficient $F_0^{(2)}$, which predicts a contribution to the scalar charge proportional to $(GM)^2\ddot\Phi \sim \mathcal O\bm( (GM\w)^2 \bm)$, cannot be determined at present.

We can now deduce the following by induction: Power counting indicates that the coefficient $F_\ell^{(n)}$ is responsible for effects appearing at order $(GM/\mathfrak R)^\ell (GM\w)^{n}$ at the earliest. Being accurate only to first order in $GM\w$, the limitations of our results in Sec.~\ref{sec:kg} preclude determining the values for any coefficient with $n\geq 2$. The $n=1$ coefficients have a one-to-one mapping with the objects $C_{\ell m}|_{a=0}$ in Eq.~\eqref{eq:kg_Clm_spherical}, so can all be determined, up to corrections in $GM\mu$, by following the same procedure that led to Eqs.~\eqref{eq:eft_matching_F01} and \eqref{eq:eft_matching_F11}.

For the $n=0$ coefficients, the vanishing of $C_{\ell m}|_{a=0}$ in the static limit $\w \to 0$ implies
\begin{equation}
\label{eq:eft_matching_F_static}
F_\ell^{(0)} \simeq 0 \quad\forall\;\ell,
\end{equation}
up to possible corrections quadratic in $GM\mu$. These coefficients are the scalar analog of a black hole's tidal Love numbers, and Eq.~\eqref{eq:eft_matching_F_static} implies that they vanish identically when $\mu=0$. (The same result is obtained in Ref.~\cite{Kol:2011vg} by different means.) It is well known that the (gravitational) tidal Love numbers also vanish~\cite{Kol:2011vg,Damour:2009vw,Binnington:2009bb,Chakrabarti:2013xza,Chakrabarti:2013lua}, which in the EFT translates to the vanishing of analogous Wilsonian coefficients that couple the black hole to the curvature tensors. This presents a fine-tuning problem, as there is no apparent symmetry in the EFT that would make this vanishing technically natural \cite{Rothstein:2014sra,Porto:2016zng}. A potential resolution has recently been put forward \cite{Penna:2018gfx}, but for now we will just accept Eq.~\eqref{eq:eft_matching_F_static} at face value. (Note that for scalars, this problem is unrelated to the no-hair theorems, which only tell us that there are no permanent scalar multipole moments; $\avg{q^L} = 0$.)

\begin{figure*}
\centering\includegraphics[width=\textwidth]{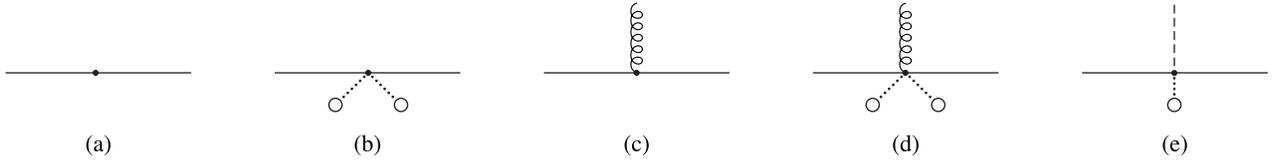}
\caption{Examples of worldline vertices. The graviton $h$ is drawn as a helical line, the scalar $\varphi$ is drawn as a dashed line, and each insertion of the background scalar $\Phi$ is denoted by a dotted line terminating in a circle. The black hole worldline, which is held nondynamical while the fields are being integrated out, is depicted as a solid line. The physical interpretation for each vertex is as follows: 
(a)~kinetic term for the black hole leading to the geodesic equation, 
(b)~correction to the kinetic term due to accretion of the background scalar, 
(c)~a graviton sourced by a black hole of constant mass, 
(d)~correction to the graviton vertex due to mass growth by accretion, and 
(e)~induced scalar charge of the black hole.}
\label{fig:worldline_vertices}
\end{figure*}

\subsection{Worldline vertices}
\label{sec:eft_worldline_vertices}

When working to leading, nontrivial order in the separation-of-scale parameters, it suffices to keep only the $F_0^{(1)}$ coefficient. At this order, the retarded propagator for $q(\tau)$ is simply
\begin{equation}
\chi_R(\tau,\tau') = \A\int_\w i\w e^{-i\w(\tau-\tau')}
\label{eq:eft_worldline_vertices_chi_R},
\end{equation}
while its commutator $\chi_C$ is just twice that. In fact, when written in this way, Eq.~\eqref{eq:eft_worldline_vertices_chi_R} is valid not only for spherical black holes, but for rotating ones as well.

We conclude this section by substituting Eq.~\eqref{eq:eft_worldline_vertices_chi_R} back into the point-particle action $S_p$ to obtain simplified expressions for the worldline vertices. This process will also help elucidate the rich physical content currently hidden in the correlation functions $\X^{AA'}(x,x')$. We begin by decomposing the fields according to Eq.~\eqref{eq:eft_matching_field_decomposition} to obtain the expansion
\begin{equation}
S_p = \sum_{n_h=0}^\infty \sum_{n_\varphi = 0}^\infty S_p^{(n_h,n_\varphi)},
\end{equation}
where the integers $(n_h,n_\varphi)$ count the number of field perturbations appearing in each term. Diagrammatic representations for the first few in this series are drawn in Fig.~\ref{fig:worldline_vertices}.

\subsubsection{Scalar terms}
\label{sec:eft_worldline_vertices_scalar}

The scalar field enters the point-particle action only through the second term in Eq.~\eqref{eq:eft_integration_dynamical_worldlines_Sp_final}. Decomposing $\phi$ according to Eq.~\eqref{eq:eft_matching_field_decomposition}, it becomes
\begin{align}
\frac{1}{2} \int_{x,x'} \left(\X^{++}\Phi\Phi'
+ 2\X^{A+}\varphi_A\Phi'
+ \X^{AA'}\varphi_A\varphi_{A'} \right),
\label{eq:eft_worldline_vertices_split_scalar}
\end{align}
having used Eq.~\eqref{eq:def_X_identity} to simplify the second term, and writing $\Phi'\equiv\Phi(x')$ for brevity. Note that the full metric is still hiding in $\X^{AA'}$, so this can be further expanded to generate an infinite series of terms with $n_h \geq 0$ and $0 \leq n_\varphi \leq 2$. Here, we concentrate on terms that depend only on~$\varphi$.

A discussion of the first term in Eq.~\eqref{eq:eft_worldline_vertices_split_scalar} is postponed until Sec.~\ref{sec:motion}. The second term, linear in $\varphi$ and drawn in Fig.~\ref{fig:worldline_vertices}(e), sources the induced scalar charge and can be rewritten as
\begin{equation}
S_p^{(0,1)} = \int_x \sqrt{-g} \Q^A(x) \varphi_A(x)
\end{equation}
upon defining the induced charge density of the black hole,
\begin{equation}
\Q^A(x) \coloneq \frac{1}{\sqrt{-g}}\int_{x'} \X^{A+}(x,x')\Phi(x').
\end{equation}
The reader will not be surprised to learn that, in the physical limit,
\begin{equation}
\label{eq:eft_worldline_vertices_scalar_PL}
\Q_+(x)|_\text{PL} 
= \int_\tau \frac{\delta^{(4)}\bm( x-z(\tau) \bm)}{\sqrt{-g}} Q(\tau), \quad
\Q_-(x)|_\text{PL} = 0.
\end{equation}
This result is derived in Appendix~\ref{sec:rules_worldline}.

\subsubsection{Graviton terms}
\label{sec:eft_worldline_vertices_gravitons}

Two terms appear in the point-particle action that are linear in the graviton $h$. In both cases, they emerge from having expanded the metric appearing in the definition of the proper time, $\dot\tau_a(g+h/\mpl) = \dot\tau_a(g) + \delta\dot\tau_a + \mathcal O(h^2)$. The first-order piece is
\begin{equation}
\label{eq:eft_worldline_vertices_delta_tau_h}
\delta\dot\tau_1(\lambda) = - \frac{1}{2\mpl}\int_x\sqrt{-g} 
h_{1,\mu\nu}(x) t^{\mu\nu}_1(x;\lambda),
\end{equation}
with a similar expression holding for $\delta\dot\tau_2$ after relabeling ${1\leftrightarrow 2}$. Writing $g(\dot z_1,\dot z_1) 
= g_{\mu\nu}(z_1) \dot z{}_1^\mu \dot z{}_1^\nu$ as shorthand,
\begin{equation}
t^{\mu\nu}_1(x;\lambda) = \frac{\dot z{}_1^\mu \dot z{}_1^\nu}
{\sqrt{-g(\dot z_1,\dot z_1)}} \frac{\delta^{(4)}\bm( x-z_1(\lambda) \bm)}{\sqrt{-g}}
\end{equation}
is the contribution to the energy-momentum tensor of a unit point mass when it is at the position $\lambda$ along the worldline $z_1$. The total energy-momentum tensor of a point mass $M$ is then obtained by simply integrating over the worldline:
\begin{equation}
T^{\mu\nu}_a(x) = M\int_\lambda t^{\mu\nu}_a(x;\lambda).
\end{equation}
Substituting this expansion into the point-mass term $- M \int c^a \dot\tau_a \subset S_p$, we get the familiar contribution
\begin{equation}
S_p^{(1,0)} \supset \frac{1}{2\mpl} \int_x \sqrt{-g} h^a_{\mu\nu}(x) T^{\mu\nu}_a(x).
\label{eq:eft_worldline_vertices_graviton}
\end{equation}
This vertex is drawn in Fig.~\ref{fig:worldline_vertices}(c). Even without explicit calculation, we know that this term sources the gravitational potential $\sim GM/r$ of the black hole.

The second contribution to $S_p^{(1,0)}$ comes from the term
\begin{equation}
\label{eq:eft_worldline_vertices_scalar_for_h}
\int_{x,x'} \X^{++}(x,x')\Phi(x)\Phi(x').
\end{equation}
To unpack this, substitute in Eq.~\eqref{eq:eft_X_explicit_++} and integrate over the delta functions contained in $\Delta_a$. Most of the terms will vanish, since $\chi_R$ is purely dissipative at leading order, so is therefore odd under time reversal. By definition, $\chi_C$ is also odd under time reversal. One therefore finds that the only nontrivial part of Eq.~\eqref{eq:eft_worldline_vertices_scalar_for_h} is
\begin{align}
\frac{1}{2} \int_{\lambda,\lambda'} \dot\tau_1\dot\tau_{2'} 
\chi_C(\tau_1,\tau_{2'})\Phi(z_1)\Phi(z_{2'}).
\end{align}
Recall, for brevity, that (un)primed indices denote functions of (un)primed variables; e.g., $\tau_1 \equiv \tau_1(\lambda)$ whereas $z_{2'} \equiv z_2(\lambda')$. At this stage, we can expand the metric entering via the proper times to first order in $h$. Technical details of this derivation are relegated to Appendix~\ref{sec:rules_worldline}. The end result is
\begin{equation}
\label{eq:eft_worldline_vertices_accretion_formula}
S_p^{(1,0)} \supset
- \int_\lambda \delta\dot\tau{}^a(\lambda)
\left[ \delta M_a(\lambda) - \delta M_a(\lambda_f) \right],
\end{equation}
where the function $\delta M_1$ is defined by
\begin{align}
\delta M_{1}(\lambda) \coloneq &\;
\A \int^\lambda_{\lambda_i} \dx\sigma \dot\Phi\bm( z_1(\sigma) \bm) 
\int_{\lambda_i}^{\lambda_f}\dx\sigma' \dot\Phi\bm( z_2(\sigma') \bm)
\nonumber\\
&\times \delta\bm( \tau_1(\sigma)-\tau_2(\sigma') \bm).
\end{align}
One obtains the definition for $\delta M_2$ by interchanging ${1\leftrightarrow 2}$. The integration limits $(\lambda_i,\lambda_f)$ appearing in these formulas are the initial and final times at which appropriate boundary conditions are specified according to the in-in formalism.

Using the expression for $\delta\dot\tau_a$ in Eq.~\eqref{eq:eft_worldline_vertices_delta_tau_h}, the first term in Eq.~\eqref{eq:eft_worldline_vertices_accretion_formula} yields
\begin{equation}
S_p^{(1,0)} \supset
\int_x \sqrt{-g} \frac{h_{a, \mu\nu}(x)}{2\mpl} \int_\lambda c^{abc} 
\delta M_b(\lambda) t_c^{\mu\nu}(x;\lambda).
\end{equation}
When compared with Eq.~\eqref{eq:eft_worldline_vertices_graviton}, we recognize that this vertex, drawn in Fig.~\ref{fig:worldline_vertices}(d), describes a graviton sourced by a black hole whose mass is slowly growing due to accretion of the background scalar. Indeed, in the physical limit, the increase in mass as a function of the proper time is
\begin{equation}
\label{eq:eft_worldline_vertices_dM_PL}
\delta M(\tau) = \delta M_+ |_\text{PL} =
\A \int_{\tau(\lambda_i)}^\tau\dx\tau' \dot\Phi{}^2\bm(z(\tau')\bm),
\end{equation}
which is exactly what we would predict from the full theory by calculating the flux of the scalar across the horizon~\cite{Jacobson:1999vr,UrenaLopez:2002du,Gregory:2017sor,Gregory:2018ghc}. What is remarkable here is that we did not put this result in by hand. After performing matching calculations to reproduce the correct behavior of the scalar charge, our EFT immediately gives us the correct accretion rate for free. This is proof that our formalism is working correctly and, more importantly, that the physics governing these two effects are one and the same. Indeed, their magnitudes are both set by the same Wilsonian coefficient $F_0^{(1)} = 16\pi$. Interestingly, this coefficient manifests as a scalar charge when it appears in the retarded propagator $\chi_R$ but is responsible for setting the accretion rate when appearing in the commutator $\chi_C$. In this light, the relation between a black hole's scalar charge and its accretion rate can be viewed as a special case of the fluctuation-dissipation theorem.

What about the second term in Eq.~\eqref{eq:eft_worldline_vertices_accretion_formula}? It is a constant contribution to the black hole mass, but one that generically diverges in the limit $\lambda_f \to \infty$. Physically, this IR divergence is signaling the breakdown of our EFT at late times. This makes intuitive sense, since an increase in the black hole's mass must be compensated for by a depletion of the surrounding scalar-field environment. Eventually, the black hole will grow to be nearly as massive as its dwindling environment, at which point there is no longer a good separation of scales. Accordingly, we should only trust this EFT for a limited duration of time. Within its period of validity, it is safe to just absorb $\delta M_+(\lambda_f)$ into a renormalization of the constant $M$ appearing in the Lagrangian, such that $M$ represents the mass of the black hole at the point when initial conditions are specified.

Another way to see that our EFT cannot be valid for all times is to differentiate Eq.~\eqref{eq:eft_worldline_vertices_dM_PL} to obtain the accretion rate
\begin{equation}
\delta\dot M(\tau) = \A \dot\Phi{}^2\bm( z(\tau) \bm).
\label{eq:eft_worldline_vertices_dM_dot}
\end{equation}
Notice that the horizon area $\A$ appearing on the rhs is that defined at some fixed time. This is only a good approximation provided $\delta M \ll M$. A more precise formula would see the constant $\A$ replaced by the instantaneous area $\A(\tau)$, but doing it properly would require a resummation involving higher-order terms. It will be interesting to explore how to do so in the future, but in practice we expect typical scalar-field environments to be dilute enough that Eq.~\eqref{eq:eft_worldline_vertices_dM_dot} is a valid approximation for long enough periods of time.


\section{Worldline dynamics}
\label{sec:motion}

Having successfully constructed our effective action, we now wish to understand its phenomenological implications. Two classes of observables are worth calculating in this theory: field expectation values, which tell us about gravitational and scalar radiation, and the equation of motion for the worldline. The general method for computing the former has already been discussed in Sec.~\ref{sec:eft_matching}. For instance, Eq.~\eqref{eq:eft_matching_Wilsonian_F01_GR} can be used to determine the profile of scalar waves (at leading order) radiated by a black hole traveling along some worldline $z^\mu(\tau)$.

To determine the trajectory of this worldline, we integrate out the bulk fields to obtain a new effective action \cite{Galley:2008ih,Galley:2009px,Galley:2010xn}
\eqskip
\begin{align}
\Gamma[z_\pm] &= -i\log\int Dh_\pm D\varphi_\pm \exp(iS)
\nonumber\\
&= S_p^{(0,0)} +
\begin{pmatrix}
\text{sum of connected} \\
\text{Feynman diagrams}
\end{pmatrix}.
\label{eq:motion_worldline_action}
\end{align}
Its equation of motion is then obtained from the extremization condition
\begin{equation}
\label{eq:motion_fd_eom}
\left.\frac{\delta\Gamma}{\delta z_-^\mu}\right|_\text{PL} = 0.
\end{equation}

The sum of Feynman diagrams in Eq.~\eqref{eq:motion_worldline_action} stems from the backreaction of the black hole onto the background fields, leading to a number of self-force effects including radiation reaction from the emission of gravitational and scalar waves. If present, interactions with other compact objects would also appear in this sum. We believe there is little to be gained from discussing these terms in generality here. Rather, they are better understood through examples and so are left to be explored further in Sec.~\ref{sec:bin}.

In this section, we concentrate on the part of the equation of motion for the worldline arising from $S_p^{(0,0)} \subset \Gamma$, which applies universally to black holes embedded in any scalar-field environment. This part of the action reads
\begin{equation}
\label{eq:motion_eom_Sp00}
S_p^{(0,0)} = -M\int_\lambda\ c^a\dot\tau_a 
+ \frac{1}{2}\int_{x,x'} \X^{++}(x,x')\Phi(x) \Phi(x').
\end{equation}
The two terms are drawn in Figs.~\ref{fig:worldline_vertices}(a) and \ref{fig:worldline_vertices}(b), respectively. Note that this action is a functional of $z_+ \coloneq (z_1 + z_2)/2$ and $z_- \coloneq z_1 - z_2$, which give the average and difference of the coordinates of the two worldline copies $(z_1,z_2)$, but do not themselves correspond to worldlines. Of course, the average coordinate tends to a description of the physical worldline, $z_+|_\text{PL} = z$, whereas $z_-|_\text{PL}=0$. The latter suggests that we can easily solve Eq.~\eqref{eq:motion_fd_eom} by Taylor expanding the action in powers of $z_-$ and reading off the linear coefficient.

Performing this expansion for $\dot\tau_1$ (note $z_1 = z_+ + z_-/2$), we obtain $\dot\tau_1(z_1) = \dot\tau_1(z) + \delta\dot\tau_1 + \mathcal O(z_-^2)$, where
\begin{equation}
\label{eq:motion_eom_d_tau_1}
\delta\dot\tau_1 =
\frac{1}{2}\left( a_\mu z_-^\mu - \frac{\dx}{\dx\tau}(u_\mu z_-^\mu) \right),
\end{equation} 
with $a^\mu \coloneq u^\alpha\nabla_\alpha u^\mu$ denoting the acceleration of the worldline. Being interested only in the physical limit, we have already taken the liberty of sending $z_+ \to z$ and parametrizing it by the proper time $\tau$. The result for $\delta\dot\tau_2$ is similar up to the change of sign $z_- \to - z_-$. Using this expansion, the point-mass term in the action gives
\begin{equation}
\label{eq:motion_eom_geodesic}
- M \int_\lambda c^a \dot\tau_a = -M\int_\tau a_\mu z_-^\mu + \mathcal O(z_-^2).
\end{equation}

As for the second term in Eq.~\eqref{eq:motion_eom_Sp00}, we demonstrated in Sec.~\ref{sec:eft_worldline_vertices_gravitons} that it simplifies to 
\begin{gather}
\frac{1}{2} \int_{\lambda,\lambda'} \dot\tau_1\dot\tau_{2'} \chi_C(\tau_1,\tau_{2'})\Phi(z_1)\Phi(z_{2'}).
\label{eq:motion_eom_accretion_raw}
\end{gather}
We now have to expand this to first order in $z_-$. There are two routes from which $z_-$ emerges: from expanding the proper times $\dot\tau \to \dot\tau + \delta\dot\tau$ and from expanding the arguments of the background scalar $\Phi$. The method for performing the first of these expansions has already been established, with the final result given in Eq.~\eqref{eq:eft_worldline_vertices_accretion_formula}. After renormalizing the IR-divergent part, we find
\begin{equation}
\label{eq:motion_eom_accretion}
S_p^{(0,0)} \supset - \int_\lambda \delta\dot\tau{}^a\delta M_a 
= -\int_\tau (\delta M a_\mu - \delta\dot M u_\mu) z_-^\mu.
\end{equation}
Second, we expand the arguments of $\Phi$ and use the antisymmetry property of $\chi_C$ to obtain
\begin{align}
S_p^{(0,0)} &\supset
-\frac{1}{2}\int_{\tau,\tau'} \chi_C(\tau,\tau') \Phi\bm( z(\tau) \bm) 
\partial_\mu\Phi\bm( z(\tau') \bm) z_-^\mu
\nonumber\\
&=
\int_\tau Q(\tau) \partial_\mu\Phi\bm( z(\tau) \bm) z_-^\mu,
\label{eq:motion_eom_fifth_force}
\end{align}
where the second line follows after writing $\chi_C$ in Fourier space and then integrating by parts.

Combining the results in Eqs.~\eqref{eq:motion_eom_geodesic}, \eqref{eq:motion_eom_accretion}, and \eqref{eq:motion_eom_fifth_force}, we learn that the equation of motion for the worldline (neglecting backreaction effects) is
\begin{equation}
\label{eq:motion_eom_final}
[M + \delta M(\tau)] a^\mu 
= -\delta\dot M(\tau) u^\mu + Q(\tau) g^{\mu\nu}\partial_\nu\Phi.
\end{equation}
The terms involving $\delta M$ administer a drag force on the black hole due to accretion, whereas the remaining term involving a derivative on $\Phi$ must be interpreted as a scalar fifth force. The reader familiar with scalar-tensor theories will find this last term a little odd, seeing as the fifth force usually appears in the equation of motion as $Q(g^{\mu\nu} + u^\mu u^\nu)\partial_\nu\Phi$ \cite{Fujii:2003pa}. In fact, we can easily put Eq.~\eqref{eq:motion_eom_final} into such a form since, by definition,
\begin{equation}
\delta\dot M = \A \dot\Phi{}^2 = -Q\dot\Phi = -Q u^\nu\partial_\nu\Phi.
\end{equation}
Thus, an equivalent way of writing Eq.~\eqref{eq:motion_eom_final} is
\begin{equation}
\label{eq:motion_eom_final_Q}
[M + \delta M(\tau)] a^\mu = Q(\tau) (g^{\mu\nu} + u^\mu u^\nu)\partial_\nu\Phi.
\end{equation}
In Sec.~\ref{sec:eft_worldline_vertices_gravitons}, we saw that the physics of the scalar charge and of accretion were one and the same, having emerged from the same term in the point-particle action. Here, this connection is made manifest at the level of the equations of motion: The scalar fifth force due to this charge includes the drag force due to accretion. It is impossible for one to exist without the other.


\section{Binary black holes\protect\\in fuzzy dark matter halos}
\label{sec:bin}

We have so far been limited in our discussion to the general features of our EFT, which apply universally to black holes embedded in any scalar-field environment. There is further insight to be gleaned from specializing to concrete systems. To complete this paper, we explore one such example involving a black hole binary embedded in a galactic fuzzy dark matter (FDM) halo. While the calculations in this section apply to astrophysical black holes of any size, our focus will center on supermassive black holes, for which effects stemming from the scalar charge $Q$ are the largest, since $Q \propto \A$.

Galactic halos in FDM models consist of a central (pseudo)solitonic core that is surrounded by an envelope of fluctuating density granules arising from wave interference \cite{Schive:2014dra,Schive:2014hza,Schwabe:2016rze,Veltmaat:2016rxo}. The core resists further gravitational collapse by coherently oscillating in time at a frequency $\w$ that is essentially set by the scalar's mass, $\w \approx \mu$,\footnote{This relation holds up to small, negative corrections from a nonrelativistic binding energy \cite{Feinblum:1968nwc,Jetzer:1991jr,PhysRevLett.66.1659,Liebling:2012fv,Visinelli:2017ooc}, which we neglect.} and has a typical length scale $\mathfrak R$ determined by the scalar's de~Broglie wavelength,
\begin{equation*}
\label{eq:bin_smbh_est_de_Broglie}
\mathfrak R
\sim 400~\text{pc}\,
\bigg( \frac{\mu}{10^{-22}~\text{eV}} \bigg)^{-1}
\bigg( \frac{v_\text{vir}}{300~\text{km}\,\text{s}^{-1}} \bigg)^{-1},
\end{equation*}
where $v_\text{vir}$ denotes the virial velocity of the halo.

As galaxies merge, the black holes at their centers form a binary that inspirals for eons before ultimately coalescing \cite{Yu2002}. In this section, we use our EFT to determine how the binary's early inspiral is affected when situated inside an FDM halo's core.\footnote{Binaries outside the core may find their orbital inspiral stalled at kiloparsec scales due to interactions with FDM fluctuations, which pump energy into the orbit \cite{Hui:2016ltb,2018arXiv180907673B}.} For simplicity, we will focus exclusively on systems for which the orbital separation $a$ is much smaller than the typical length scale $\mathfrak R$ of the background. Even a gargantuan $10^{10}~M_\odot$ black hole has a radius that extends only to a few milliparsecs; thus, it is easy to envision comfortably fitting not just one black hole, but a binary of supermassive black holes within such a distance. Calculations are straightforward in this regime because the constituents of the binary perceive a local environment that is effectively spatially homogeneous:
\begin{equation}
\label{eq:bin_smbh_bg}
\Phi = \Phi_0 \cos (\mu t + \Upsilon) + \mathcal O(a/\mathfrak R),
\end{equation}
where $\Upsilon$ is some arbitrary phase. Let ${\eps = \Phi_0/\mpl}$ be a dimensionless parameter that characterizes the local density of this halo. Typical FDM halos satisfy the condition $\eps\ll 1$ [see also Eq.~\eqref{eq:bin_smbh_order_estimate} later]; hence, the scalar field backreacts onto the geometry only weakly. As a result, we can expand the background metric as $g = \eta + H$ about Minkowski space, where $H \sim \mathcal O(\eps^2)$ is the gravitational potential of the halo.

Provided that background gradients $\partial H$ are not too strong (a more precise statement will be made in Sec.~\ref{sec:bin_dynamics_background_potential}), the dominant force acting on the black holes is still their mutual gravitational attraction. In such circumstances, the virial theorem relates the orbital separation of the binary to the typical size $GM$ and the characteristic velocity $v$ of its constituents; $v^2 \sim GM/a$. For most of its inspiral, ${v \ll 1}$, allowing us to study the evolution of this system in the nonrelativistic, post-Newtonian (PN) limit.

Furthermore, when $v$ is small, the system neatly separates into a ``near zone'' and a ``far zone.'' Following Refs.~\cite{Goldberger:2004jt,Huang:2018pbu,Kuntz:2019zef}, these two zones are dealt with one at a time by constructing a tower of EFTs. To that end, we split
\begin{equation*}
(h,\varphi) \to (\bar h,\bar\varphi) + (h,\varphi).
\end{equation*}
The fields $(h,\varphi)$ on the rhs represent potential modes that mediate forces in the near zone (at distances $r\sim a$), whereas $(\bar h,\bar\varphi)$ denote radiation modes in the far zone ($r \gtrsim a/v$). The former are always off shell, whereas the latter can go on shell and propagate to infinity. The potential modes are integrated out first to obtain a new effective action governing the dynamics of the binary coupled to the remaining radiative \dof,
\begin{equation}
S_\text{eff}[z_\K,\bar h,\bar\varphi] = -i\log \int D h_\pm D \varphi_\pm \exp (iS),
\end{equation}
where $S = S_f + \sum_\K S_{p,\K}$ is the original gauge-fixed\footnote{We gauge fix the potential mode $h$ with respect to the background $g = \eta + H + \bar h/\mpl$ to preserve gauge invariance of $S_\text{eff}$ \cite{Goldberger:2004jt}.} effective action [cf.~Eq.~\eqref{eq:eft_S}] and the index $\K \in \{1,2\}$ labels the individual members of the binary. The flux of radiative modes off to infinity can be calculated at this stage using~$S_\text{eff}$. The effective action for the self-consistent motion of the worldlines is obtained after also integrating out the radiation modes [cf.~Eq.~\eqref{eq:motion_worldline_action}]:
\begin{equation}
\Gamma[z_\K] = -i\log \int D\bar h_\pm D\bar\varphi_\pm \exp (iS_\text{eff}).
\end{equation}

A convenient way to perform these integrations in perturbation theory is with the use of Feynman diagrams, which can be organized to scale in a definite way with the expansion parameters of our EFT. Schematically, each term in the effective action $\Gamma$ scales as
\begin{equation*}
\Gamma \sim L^{1-\ell} \, v^{2n} \, \eps^{p_1} \, (GM\mu)^{p_2},
\end{equation*}
where $L \sim Mav$ is the characteristic angular momentum of the binary. The integer $\ell$ counts the number of loops in a given Feynman diagram, and since $L \gg 1$ for astrophysical black holes, only the tree-level contributions are needed \cite{Goldberger:2004jt}. The integer or half-integer $n$ counts the order in the usual PN expansion, which is supplemented by two additional parameters, $\eps$ and $GM\mu$, that characterize the impact of the scalar-field environment on the binary.\footnote{Of the three separation-of-scale parameters we started with in Eq.~\eqref{eq:eft_small_parameters}, only $GM\mu$ survives because we neglect spatial variations of $\Phi$ and have set $\w \approx \mu$.} The terms with $p_1 = p_2 = 0$ constitute the standard PN equations for a binary in vacuum \cite{Blanchet:2013haa}, and need not be revisited here. Effects involving the scalar field first appear when $p_1 = p_2 = 2$. As in earlier parts of this paper, we work only at leading nontrivial order; hence, our EFT is, in fact, organized as an expansion in just two small parameters: $v$ and $\eps GM\mu$.

\subsection{Phenomenology}
\label{sec:bin_dynamics}

In what follows, we discuss five distinct physical effects that arise when a black hole binary is embedded in an FDM halo. Concomitantly with some explicit calculations, we also establish power counting rules to determine the order at which they appear in the PN expansion. As the effects we discuss span a range of 4.5PN orders, a comprehensive and systematic expansion of $\Gamma$ in powers of $v$ is far beyond the scope of this paper. We will limit ourselves to deriving only the leading-order expression for each effect.

\subsubsection{Scalar dipole radiation}
\label{sec:bin_dynamics_radiation}

It is only fitting that we begin our discussion with the phenomenon that started it all. In the PN limit, the coordinate time~$t$ can be used to parametrize the worldlines; hence, the charge densities in Sec.~\ref{sec:eft_worldline_vertices_scalar} reduce to
\begin{equation}
\mathcal Q^A_\K(x) = Q_\K(t) \delta^A_\K(x)
\end{equation}
at leading order in $v$, where the delta function
\begin{equation}
\delta_{a,\K}(x) \coloneq \delta^{(3)}\bm( \bmf x - \bmf z_{a,\K}(t) \bm)
\end{equation}
localizes the integral to be along the $a$th copy of the $\K$th worldline. For the background in Eq.~\eqref{eq:bin_smbh_bg}, the scalar charge is $Q_\K(t) = \A_\K \mu\Phi_0 \sin(\mu t + \Upsilon)$. The radiation mode $\bar\varphi$ couples to the binary via the term
\begin{equation}
S_\text{eff} \supset \sum_\K \int_x \mathcal Q^A_\K(x) \bar\varphi_A(x).
\label{eq:bin_smbh_action_rad}
\end{equation}

Definite scaling in $v$ is achieved by multipole expanding the radiation mode as~\cite{Goldberger:2004jt}
\begin{equation}
\label{eq:bin_multipole_expansion_scalar}
\bar\varphi_A(t,\bmf x) = \bar\varphi_A(t,\bmf 0) + \bmf x^i \partial_i \bar\varphi_A(t,\bmf 0) + \cdots
\end{equation}
about the binary's barycenter, which we place at the origin. Substituting this back into Eq.~\eqref{eq:bin_smbh_action_rad}, we find that the monopole term $\propto \bar\varphi_A(t,\bmf 0)$ does not radiate at this PN order but merely describes the total scalar charge of the binary. The dominant channel for scalar radiation is the dipole moment, whose term in the action reads
\begin{equation}
\label{eq:bin_smbh_action_rad_dipole}
\sum_\K\int_x \mathcal Q^A_\K(x) \bmf x^i \partial_i \bar\varphi_A(x).
\end{equation}
In the physical limit, this leads to the expectation value
\begin{equation}
\avg{\bar\varphi(x)} \supset - \frac{\partial}{\partial \bmf x^i}
\int_{t'} G_R(t,\bmf x; t',\bmf 0) \bmf P^i(t'),
\label{eq:bin_smbh_rad_exp_value}
\end{equation}
sourced by the binary's scalar dipole moment
\begin{equation}
\label{eq:bin_smbh_dipole_moment}
\bmf P^i(t) = \sum_\K Q_\K(t) \bmf z_\K^i(t).	
\end{equation}

\begin{table*}
\caption{Post-Newtonian power counting rules for black hole binaries embedded in fuzzy dark matter halos. All derivatives $\partial_\mu$ scale in the same way, except spatial derivatives acting on the potential modes, which are denoted by the 3-momentum $\bmf p$, and spatial derivatives on $\Phi$, which vanish. The rules involving the radiation modes assume $\Omega \gg \mu$ for simplicity.}
\label{table:power_counting_smbh}
\begin{ruledtabular}
\begin{tabular}{ccccccccc}
$h,\varphi$ &
$\bar h,\bar\varphi$ &
$\Phi/\mpl$ &
$H$ &
$\partial_\mu$ &
$\bmf p$ &
$M/\mpl$ &
$\delta M/\mpl$ &
$Q$ \\[0.3em]
\hline
$\sqrt{v}/a$ &
$v/a$ &
$\eps \mu a/v$ &
$(\eps \mu a /v)^2$ &
$v/a$ &
$1/a$ &
$\sqrt{Lv}$ &
$\sqrt{Lv} (\eps GM\mu)^2 v^{-3}$ &
$\sqrt{Lv} (\eps GM\mu)$ \\
\end{tabular}
\end{ruledtabular}
\end{table*}

The master integral in Appendix~\ref{sec:master_integral} can be used to evaluate Eq.~\eqref{eq:bin_smbh_rad_exp_value}. Keeping only the radiative part that reaches an observer at infinity, we find
\begin{equation}
\avg{\bar\varphi(x)} \supset - \frac{1}{4\pi r} 2 \Real \int_{\mu^+}^\infty \frac{\dx\w}{2\pi} \hat{\bmf x} \cdot \tilde{\bmf P}(\w)
 ik e^{-i(\w t - kr)},
\end{equation}
where the wave~number $k = \sqrt{\w^2 - \mu^2}$. Finally, we integrate the $(t,r)$~component of the scalar's energy-momentum tensor over a spherical shell of radius $r$ and discard terms that vanish in the limit $r\to\infty$ to obtain the radiated power
\begin{equation}
\mathcal F_\phi = - r^2 \int \dx^2\Omega \partial_r \avg{\bar\varphi} \partial_t \avg{\bar\varphi}.
\end{equation}
For a circular binary with orbital frequency $\Omega$, the flux at a distance $r$ is
\begin{align}
\mathcal F_\phi
=&\,
\frac{16\pi \mpl^2}{3} (\eps  GM\mu)^2 (GM\Omega)^{8/3} \nu^2
\left( \frac{M_1-M_2}{M} \right)^2
\nonumber\\ &\times
\left[
v_+^3 \frac{\Omega_+^4}{\Omega^4} + \theta(\Omega-2\mu) v_-^3 \frac{\Omega_-^4}{\Omega^4}
\right.
\nonumber\\ & 
\left.
~ - \theta(\Omega-2\mu)\, v_+ v_- (v_+ + v_-)  \frac{\Omega_+^2\Omega_-^2}{\Omega^4} \cos\varpi
\right],
\label{eq:bin_smbh_flux}
\end{align}
where $M = M_1 + M_2$ is the total mass of the binary and $\nu = M_1 M_2/M^2$ is its symmetric mass ratio.

Four worthy observations can be made here: First, the terms in square brackets signify that scalar waves emanate at two frequencies, ${\Omega_\pm = \Omega \pm \mu}$. This is to be expected since the dipole moment $\tilde{\bmf P}^i(\w)$ is the convolution of $Q_\K$ and~$\bmf z_\K$. The two waves travel with different group velocities ${v_\pm = (1-\mu^2/\Omega_\pm^2)^{1/2}}$, and the third line in Eq.~\eqref{eq:bin_smbh_flux} accounts for their interference after they accumulate a phase difference ${\varpi = 2\mu t + 2\Upsilon - (\Omega_+ v_+ - \Omega_- v_-)r}$.  Second, the presence of step functions indicates that the larger-frequency mode $\Omega_+$ is radiated throughout the entire history of the inspiral, whereas the lower-frequency mode $\Omega_-$ is radiated only when ${\Omega_- > \mu}$. This stems from the simple fact that only sources with frequencies greater than the scalar's mass can deposit energy into on-shell modes. Third, observe that the flux vanishes entirely in the equal-mass limit. We can understand this by noticing in Eq.~\eqref{eq:bin_smbh_dipole_moment} that the dipole moment becomes proportional to the position of the barycenter when $M_1 = M_2$. Finally, as a sanity check, we note that Eq.~\eqref{eq:bin_smbh_flux} reduces to the correct expression [Eq.~(2.37) of Ref.~\cite{Horbatsch:2011ye}\,] in the massless limit $\mu \to 0$ with $Q_\K \to \text{const}$.

Let us clarify when our result for $\mathcal F_\phi$ is valid. It relies on the multipole expansion in Eq.~\eqref{eq:bin_multipole_expansion_scalar}, which holds if the larger-frequency mode, with momentum $|\bmf p| = (\Omega_+^2 - \mu^2)^{1/2}$, satisfies $a |\bmf p| \ll 1$. Writing $a^2 \bmf p^2 = a^2 \Omega^2 + 2a^2\mu\Omega$, we can rephrase this as two conditions: We require $a^2\Omega^2 \ll 1$ and $a^2\mu\Omega \ll 1$. The first of these equivalently reads $v^2 \ll 1$, so is always satisfied during the early inspiral. The second can be rewritten as $\mu a v \ll 1$ or $a \ll 1/(GM\mu^2)$ and signifies that the binary cannot be too widely separated;
\begin{equation}
a \ll 10~\text{pc}\,
\bigg(\frac{M}{10^{10}~M_\odot} \bigg)^{-1}
\bigg(\frac{\mu}{10^{-22}~\text{eV}}\bigg)^{-2}.
\label{eq:bin_smbh_instantaneous_scalar_condition}
\end{equation}
We may regard this condition as an IR cutoff for the validity of our EFT when applied to this system.

This kind of scaling analysis can also be used to establish power counting rules, which enable a quick estimate of the relative sizes of different effects. (The rules developed here and later in this section are summarized in Table~\ref{table:power_counting_smbh}.) For simplicity, we will concentrate on the later stages of the inspiral (${\Omega \gg \mu}$) when discussing radiative effects, since this is when they are most pronounced. In this regime, the 4-momentum of $\bar\varphi$ satisfies $p \sim \Omega \sim v/a$; thus, the propagator scales as $\avg{\bar\varphi\bar\varphi} \sim \int\dx^4p/p^2 \sim (v/a)^2$, and so $\bar\varphi \sim v/a$ when appearing as an internal line in a Feynman diagram. Similar reasoning implies $\bar h \sim v/a$~\cite{Goldberger:2004jt}. In position space, the 4-momentum $p_\mu$ translates into a derivative $\partial_\mu$; thus, $\partial_\mu \sim v/a$ when acting on the radiation modes. Time derivatives acting on the background scalar can be arranged to scale in the same way by taking $\Phi/\mpl \sim \eps\mu a/v$, such that $\partial_t \Phi \sim \mu \eps \mpl$. Consequently, $Q \sim \sqrt{Lv}(\eps GM\mu)$ after using the relation $M/\mpl \sim \sqrt{Lv}$~\cite{Goldberger:2004jt}. We use these rules to deduce that Eq.~\eqref{eq:bin_smbh_action_rad_dipole} scales as
\begin{equation*}
\int \dx t \mathcal Q \bmf x^i \partial_i \bar\varphi \sim
\left(\frac{a}{v}\right) Q a \left(\frac{v}{a}\right)^2 \sim \sqrt{L} v^{3/2} (\eps GM\mu),
\end{equation*}
where $\int \dx t \sim a/v$, since the orbital period is the key timescale in this system. Integrating out the radiation modes, two copies of this vertex linked by a propagator generate a term in $\Gamma$ that scales as $L v^3 (\eps GM\mu)^2$. Hence, scalar radiation reaction first appears at 1.5PN order, albeit suppressed by two powers of $\eps GM\mu$. For typical FDM halos \cite{Schive:2014dra,Schive:2014hza,Bar:2018acw,Boskovic:2018rub},
\begin{equation}
(\eps GM \mu)^2 \sim 2 \times 10^{-16}\,
\bigg(\frac{\rho}{100~M_\odot\,\text{pc}^{-3}}\bigg)
\bigg(\frac{M}{10^{10}~M_\odot}\bigg)^2,
\label{eq:bin_smbh_order_estimate}
\end{equation}
where $\rho = \mu^2 \Phi_0^2 / 2$ is the local energy density.

It is instructive to compare this with gravitational radiation reaction, which scales as $L v^5$ (2.5PN order). Our power counting rules then tell us that the energy radiated in scalar waves is suppressed by $(\eps GM\mu )^2 /v^2$ relative to gravitational waves. Indeed, this simple estimate is consistent with a more detailed calculation. In the ${\Omega \gg \mu}$ limit, the ratio of $\mathcal F_\phi$ to the leading quadrupolar flux of gravitational waves $\mathcal F_g$ \cite{Blanchet:2013haa} is
\begin{equation}
\frac{\mathcal F_\phi}{\mathcal F_g}
\sim
\frac{5}{48} \frac{(\eps GM\mu)^2}{v^2}\left( \frac{M_1 - M_2}{M} \right)^2 \sin^2(\mu t + \Upsilon).
\end{equation}
Taking $v\approx 0.1$ and $(\eps GM\mu)^2 \approx 2 \times 10^{-16}$, this ratio is at most $2\times 10^{-15}$. Clearly, the impact of scalar radiation on the inspiral of the binary is unlikely to be observable. That said, effects appearing at lower PN orders may have better observational prospects; hence, the remainder of this section concentrates on terms in $\Gamma$ that arise from integrating out the potential modes.

\subsubsection{Scalar fifth force}
\label{sec:bin_dynamics_fifth_force}

\begin{figure}
\centering\includegraphics{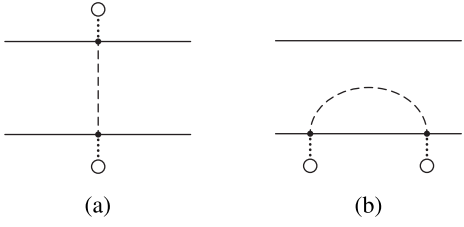}
\caption{(a) The exchange of a potential-mode scalar between the worldlines mediates an attractive scalar fifth force. (b) Self-energy diagram that is pure counterterm. Its mirror inverse, in which the scalar propagates to and from the top worldline, is included implicitly since we do not distinguish between the two solid lines.}
\label{fig:bin_smbh_fifth_force}
\end{figure}

The potential mode $\varphi$ couples to the scalar charge of the black hole in the same way as $\bar\varphi$, namely through the term
\begin{equation}
S_{p,\K} \supset \int_x \mathcal Q_\K^A(x) \varphi_A(x).
\end{equation}
The diagrams in Fig.~\ref{fig:bin_smbh_fifth_force} arise from connecting two copies of this vertex by a propagator. Using standard Feynman rules (outlined in Appendix~\ref{sec:rules_bulk}), they yield
\begin{align}
\text{Fig.~\ref*{fig:bin_smbh_fifth_force}}
=
\sum_{\K,\K'}\int_{x,x'} Q_\K(t) \delta_\K^+(x) G_R(x,x') \delta_{\K'}^-(x') Q_{\K'}(t').
\end{align}
The sum over terms with ${\K = \K'}$ leads to the self-energy diagram in Fig.~\ref{fig:bin_smbh_fifth_force}(b), which is pure counterterm and vanishes identically in dimensional regularization \cite{Goldberger:2004jt} (at leading order in $GM\mu$). Only the cross terms ${\K \neq \K'}$ in Fig.~\ref{fig:bin_smbh_fifth_force}(a) have interesting physical consequences. As we did in Sec.~\ref{sec:motion}, the equations of motion for the worldlines can be read off after expanding each term in $\Gamma$ to first order in $\bmf z_-$. We use the fact that
\begin{subequations}
\label{eq:bin_smbh_delta_function_expand}
\begin{align}
\delta_\K^+(x) &= - \bmf z_-^i \frac{\partial}{\partial\bmf x^i} \delta^{(3)}\bm( \bmf x - \bmf z_{\K}(t) \bm) + \mathcal O(\bmf z_-^2),
\\
\delta_\K^-(x) &= \delta^{(3)}\bm( \bmf x - \bmf z_{\K}(t) \bm) + \mathcal O(\bmf z_-)
\end{align}
\end{subequations}
to write
\begin{gather}
\text{Fig.~\ref*{fig:bin_smbh_fifth_force}(a)}
=
\sum_{\K\neq \K'} \int_{t,t'} \frac{\partial}{\partial\bmf z_\K^i}
\left[
 Q_K Q_{\K'} G_R(t,\bmf z_\K; t',\bmf z_{\K'})
 \right] \bmf z_{-,\K}^i.
\label{eq:bin_smbh_fifth_force_GR}
\end{gather}

It is instructive to first evaluate this integral while holding the black holes fixed at their respective positions. This permits use of the master integral in Appendix~\ref{sec:master_integral}, which returns
\begin{equation}
\text{Fig.~\ref*{fig:bin_smbh_fifth_force}(a)} = \sum_{\K \neq \K'} \int_t \frac{\partial}{\partial\bmf z_\K^i}
\left( \frac{Q_\K Q_{\K'}}{4\pi |\bmf z_\K - \bmf z_{\K'}|} \right) \bmf z_{-,\K}^i.
\end{equation}
In general, the $\K$th black hole obeys an equation of motion of the form $M_\K \bmf a_\K = \bmf F_\K$. Taking the functional derivative of Fig.~\ref{fig:bin_smbh_fifth_force}(a) with respect to $\bmf z_{-}$, we learn that the first black hole experiences the scalar fifth force
\begin{equation}
\bmf F_1 \supset -\frac{Q_1 Q_2}{4\pi r^2} \bmf n,
\end{equation}
where ${\bmf r = \bmf z_1 - \bmf z_2}$, ${r = |\bmf r|}$, and ${\bmf n = \bmf r/r}$. Naturally, interchanging the labels $1 \leftrightarrow 2$ yields an identical force acting on the second black hole.\footnote{Note that these labels now distinguish between the members of the binary. Equations of motion are always given in the physical limit, so there are no longer any CTP indices floating around.}

We obtained this result by keeping the black holes at rest, but nothing changes at this PN order had they been allowed to move freely, since any departure from the static case must depend on $v$. When working to higher orders, we achieve definite scaling in powers of $v$ by expanding the propagator for the potential mode $\varphi$ about its instantaneous limit:
\begin{equation}
\tilde G_R(p) = 
\frac{1}{p^2 + \mu^2}
=
\frac{1}{\bmf p^2}
\left(
1 + \frac{(p^0)^2-\mu^2}{\bmf p^2} + \cdots
\right).
\label{eq:bin_smbh_propagators_instantaneous_limit_scalar}
\end{equation}
The instantaneous part $1/\bmf p^2$ is responsible for the inverse-square law force; hence, the potential mode has 3-momenta satisfying $\bmf p \sim 1/a$, or in other words, spatial derivatives acting on $\varphi$ scale as $1/a$. In contrast, the oscillating background forces the energy of the scalar to have two pieces that scale differently: $|p^0| \sim \mu + v/a$, such that $[(p^0)^2-\mu^2]/\bmf p^2 \sim v^2 + \mu a v$. Thus, we see that assuming $\varphi$ propagates instantaneously is valid only under the conditions $v^2 \ll 1$ and $\mu av \ll 1$, which are the same conditions we derived earlier for the radiation modes; cf.~Eq.~\eqref{eq:bin_smbh_instantaneous_scalar_condition}.

For the power counting rules, it suffices to neglect the subleading $\mu a v$ dependence when working to leading order in $GM\mu$,\footnote{At higher orders, it becomes necessary to factor it out explicitly; for instance, by working with the complex field $\psi$ instead, defined from $\varphi(x) \propto [e^{-i \mu t} \psi(x) + \text{H.c.}]$.} such that $(p^0)^2 - \mu^2 \sim v^2 / a^2$, while time derivatives of $\varphi$ scale with $v/a$. Taken together, these considerations imply $\varphi \sim \sqrt{v}/a$. Similar relations apply to the potential-mode graviton $h$ (see Table~\ref{table:power_counting_smbh}), whose propagator admits the analogous quasi-instantaneous expansion \cite{Goldberger:2004jt}
\begin{equation}
\tilde D_R(p) =
\frac{1}{p^2} = \frac{1}{\bmf p^2} \left( 1 + \frac{(p^0)^2}{\bmf p^2} + \cdots  \right).
\label{eq:bin_smbh_propagators_instantaneous_limit_graviton}
\end{equation}
These power counting rules tell us that $\text{Fig.~\ref{fig:bin_smbh_fifth_force}} \sim L v^0 (\eps GM\mu)^2$; thus, the scalar fifth force is a Newtonian-order effect.

\subsubsection{Accretion}
\label{sec:bin_dynamics_accretion}

\begin{figure}
\centering\includegraphics{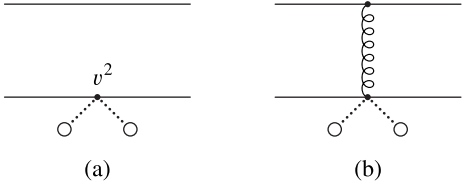}
\caption{Leading-order diagrams accounting for accretion. As with earlier diagrams, their mirror inverses (in which the background scalar interacts with the top worldline) are included implicitly.}
\label{fig:bin_smbh_accretion}
\end{figure}

We already encountered the drag force from accretion in Sec.~\ref{sec:motion} in a fully relativistic setting. When expanded in powers of $v$, the leading term in Eq.~\eqref{eq:motion_eom_accretion_raw} is proportional to $v^2$ and is depicted in Fig.~\ref{fig:bin_smbh_accretion}(a).\footnote{There would also have been an $\mathcal O(v^0)$ term if $\Phi$ were spatially inhomogeneous, which would yield the scalar fifth force $\propto \partial_i \Phi$ exerted by the background; cf. Eq.~\eqref{eq:motion_eom_final}.} Schematically,
\begin{equation*}
\text{Fig.~\ref*{fig:bin_smbh_accretion}(a)} \sim \int \dx t\, \frac{1}{2} \delta M(t) \bmf v^2 \sim
L v^{-3} (\eps GM\mu)^2.
\end{equation*}

In the presence of a second black hole, an additional diagram contributes at this order: Fig.~\ref{fig:bin_smbh_accretion}(b) accounts for the change in the gravitational force between the black holes due to their increasing masses. Notice that only one of the black holes is accreting in this diagram; the diagram in which both are accreting first appears at $\mathcal O(\eps^4)$ and, thus, is neglected.

Even without detailed calculation, it is easy to correctly intuit that Fig.~\ref{fig:bin_smbh_accretion} leads to the force
\begin{equation}
\bmf F_1 \supset - \delta\dot M_1 \bmf v_1 - \delta M_1 \bmf a_1 - \frac{G (M_1 \delta M_2 + M_2 \delta M_1)}{r^2} \bmf n.
\label{eq:bin_smbh_accretion}
\end{equation}
Formally, this is a $-1.5$PN effect but is still subleading to the Newtonian-order interactions $\sim L v^0$ due to suppression by two powers of $\eps GM \mu$. The negative-power scaling in $v$ indicates that the effects of accretion---in contrast to radiation reaction---are most pronounced at the very early stages of the inspiral when the binary is widely separated. Consequently, future space-based gravitational-wave detectors like LISA are unlikely to be sensitive to this effect. Rather, pulsar timing arrays or other astronomical observations may prove more suitable when attempting to observe, or at least constrain, the impact of an FDM halo on a supermassive black hole binary. We will return to the subject of constraining FDM models in Sec.~\ref{sec:bin_constraints}.

\subsubsection{Background gravitational potential}
\label{sec:bin_dynamics_background_potential}

The three effects discussed so far---scalar radiation, the fifth force, and accretion---all stem from the interaction between a black hole's horizon and the scalar field. Two other effects, which are not unique to black holes but which influence the motion of any massive body, can also be calculated using our EFT framework. We discuss the external force due to the halo's gravitational potential here, before turning to dynamical friction in Sec.~\ref{sec:bin_dynamics_dynamical_friction}.

As we did for the radiation modes, we preserve definite scaling in $v$ by multipole expanding
\begin{align}
H_{\mu\nu}(t,\bmf x) =&\;
H_{\mu\nu}(t,\bmf 0)
+ \bmf x^i\partial_i H_{\mu\nu}(t,\bmf 0)
\nonumber\\
&
+ \frac{1}{2} \bmf x^i \bmf x^j \partial_i\partial_j H_{\mu\nu}(t,\bmf 0) + \cdots
\label{eq:bin_smbh_H_multipole_expansion}
\end{align}
about the binary's barycenter. Note $H_{\mu\nu}$ must depend on the spatial coordinates---despite $\Phi$ being (approximately) just a function of time---if it is to be a consistent solution at $\mathcal O(\eps^2)$ to the background field equation
\begin{equation}
R_{\mu\nu} = 8\pi G \left(\partial_\mu\Phi \partial_\nu\Phi + \frac{1}{2} \eta_{\mu\nu} \mu^2\Phi^2 \right).
\end{equation}
This equation enforces the relation $R \sim \partial\partial H \sim \eps^2\mu^2$, which is satisfied provided all derivatives acting on $H$ scale as $\partial_\mu \sim v/a$, while taking $H \sim (\eps\mu a/v)^2$.

Although it is possible to stick with the general multipole expansion in Eq.~\eqref{eq:bin_smbh_H_multipole_expansion}, it is far more convenient if we work in Fermi normal coordinates \cite{MTW,Poisson2011}. We then have that both $H_{\mu\nu}(t,\bmf 0)$ and $\partial_i H_{\mu\nu}(t,\bmf 0) = 0$ in this gauge,\footnote{The linear terms $\partial_i H$ would be nontrivial if $\partial_i\Phi\neq 0$.} whereas
\begin{subequations}
\begin{align}
\frac{1}{2}\partial_i\partial_j H_{00}(t,\bmf 0) &= - R_{0i0j}(t,\bmf 0),
\\
\frac{1}{2}\partial_i\partial_j H_{0k}(t,\bmf 0) &= -\frac{2}{3} R_{0ikj}(t,\bmf 0),
\\
\frac{1}{2}\partial_i\partial_j H_{k\ell}(t,\bmf 0) &= -\frac{1}{3} R_{ki\ell j}(t,\bmf 0).
\end{align}
\end{subequations}
At leading order in the PN expansion, the only contribution involving $H_{\mu\nu}$ comes from expanding the point-mass term in the action:
\begin{align}
\Gamma &\supset
\sum_\K c^a \int_t \frac{1}{2} M_\K H_{00}\bm( t,\bmf z_{a,\K}(t) \bm)
\nonumber\\
&=
- \sum_\K \int_t M_\K R_{0i0j}(t,\bmf 0) \bmf z_{-,\K}^i \bmf z_{+,\K}^j,
\label{eq:bin_smbh_background_potential_term}
\end{align}
which gives rise to the force
\begin{equation}
\bmf F_\K^i \supset - R^i{}_{0j0}(t,\bmf 0) M_\K \bmf z_\K^j.
\end{equation}
The effect of this force on binary pulsars has previously been studied in Ref.~\cite{Blas:2016ddr}. Its effect on black hole binaries is analogous and will be discussed briefly in Sec.~\ref{sec:bin_constraints}.

Power counting tells us that this external force first appears at $-3$PN order; $\text{Eq.~\eqref{eq:bin_smbh_background_potential_term}} \sim L v^{-6} (\eps GM\mu)^2$. This inverse scaling with $v$, which we first met in Sec.~\ref{sec:bin_dynamics_accretion}, is signaling a second type of IR breakdown of our EFT.\footnote{The first type has to do with a breakdown at arbitrarily late times; see the last few paragraphs of Sec.~\ref{sec:eft_worldline_vertices_gravitons}.} To see this, recall that our perturbative expansion is predicated on the virial relation $v^2 \sim GM/a$, which holds only if the Newtonian-order interactions $\sim L v^0$ are the dominant terms in the action. This demands that the binary satisfy the condition ${v^{6} \gg (\eps GM\mu)^2}$, which can equivalently be written as ${a^3 \ll GM/(\eps\mu)^2}$ or most transparently as ${(\eps\mu a / v)^2 \sim H \ll 1}$. For small enough velocities or large enough orbital separations, our scaling rules naively suggest that $H$ can attain values of order one, at which point it stops being a weak perturbation to the Minkowski metric. Before this can happen, spatial variations of $\Phi$ become relevant and must be taken into account. Thus, a multipole expansion of the background fields is valid only if
\begin{equation}
a \ll 80\,\text{pc}\,
\bigg( \frac{\rho}{100~M_\odot\,\text{pc}^{-3}} \bigg)^{-1/3}
\bigg( \frac{M}{10^{10}~M_\odot} \bigg)^{1/3}.
\label{eq:bin_smbh_est_max_a}
\end{equation}
This is a second, independent IR cutoff for our EFT, which must be satisfied in addition to Eq.~\eqref{eq:bin_smbh_instantaneous_scalar_condition}.

\subsubsection{Dynamical friction}
\label{sec:bin_dynamics_dynamical_friction}

\begin{figure}
\centering\includegraphics{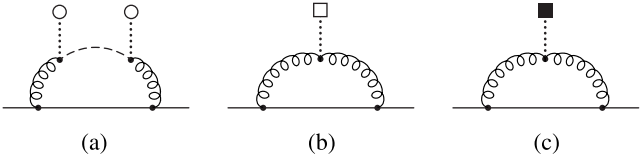}
\caption{Feynman diagrams constituting (a) dynamical friction and (b),(c) the backscattering of gravitons. Details about the interaction vertices in the bulk can be found in Appendix~\ref{sec:rules_bulk}.}
\label{fig:bin_smbh_dynamical_friction}
\end{figure}

The final effect we wish to discuss is the drag force due to dynamical friction. It arises because the gravitational field of a black hole, or any massive body, perturbs the medium through which it moves, forming a wake in the latter that then exerts a gravitational pull back on the object. Although usually considered in the context of collisionless or gaseous media~\cite{Chandrasekhar1943,BinneyTremaine,Ostriker:1998fa,Kim:2007zb}, recent studies have begun exploring what modifications are needed to account for the wavelike nature of FDM \cite{Hui:2016ltb,Lora:2011yc,2018arXiv180907673B}. Our EFT formalism provides a natural language for calculating the force that dynamical friction exerts on a massive body. The interaction of a black hole with its gravitationally induced wake is depicted in Fig.~\ref{fig:bin_smbh_dynamical_friction}(a), and yields
\begin{align}
\bmf F_\K &\supset
-16 \pi (GM_\K)^2 \dot\Phi{}^2 \bmf v_\K^2 \hat{\bmf v}_\K.
\label{eq:bin_smbh_dynamical_friction}
\end{align}
The derivation is presented in Appendix~\ref{sec:derivation_dynamical_friction}.

This formula relies on the assumption that the binary is tight enough to satisfy the condition $\mu a v \ll 1$ [cf.~Eq.~\eqref{eq:bin_smbh_instantaneous_scalar_condition}] such that the scalar can be approximated as propagating instantaneously at leading PN order. In Refs.~\cite{Lora:2011yc,Hui:2016ltb,2018arXiv180907673B}, the impact of dynamical friction within an FDM halo is studied in the opposite regime $\mu a v \gtrsim 1$ (objects orbiting the center of a galaxy, for instance, satisfy this condition). Consequently, our results cannot be directly compared and they need not agree. We have, however, verified that our EFT approach correctly reproduces the results in Appendix~A of Ref.~\cite{Lora:2011yc} when working under similar assumptions.

Let us return to our own result in Eq.~\eqref{eq:bin_smbh_dynamical_friction}: Power counting tells us that dynamical friction first appears at $-1$PN order. In contrast, the diagrams in Figs.~\ref{fig:bin_smbh_dynamical_friction}(b) and \ref{fig:bin_smbh_dynamical_friction}(c)---which depict the backscattering of gravitons off the gravitational potential and energy density of the halo---scale with $\eps$ and $GM\mu$ in the same way but appear earlier at $-2$PN order. Evaluating these diagrams proves to be more challenging, however, and is for the time being left as an open problem.

\subsection{Observational constraints}
\label{sec:bin_constraints}

We conclude this section by exploring how well observations of OJ287 can be used to constrain FDM models. The supermassive black hole binary at the center of the quasar has an orbital period that decays at a rate ${\dot P \sim 10^{-3}}$, which is consistent with the predictions of vacuum PN theory to within an uncertainty of 6\% \cite{Dey:2018mjg}. Hence, the effects discussed in Sec.~\ref{sec:bin_dynamics} should not hasten or stall the inspiral by more than $|\delta \dot P| = 6 \times 10^{-5}$. This condition can be translated into an upper bound on the local FDM density $\rho$ in the vicinity of the quasar. Although PN corrections are needed to accurately predict the evolution of the inspiral due to gravitational-wave emission \cite{Dey:2018mjg}, it suffices to treat effects involving the scalar field as first-order perturbations to the Kepler problem when determining their contribution to $\dot P$. The general method for performing such calculations is described at length in Chap.~3 of Poisson and Will \cite{PoissonWill}; in what follows, we will simply quote the required formulas.

Consider the effective-one-body Kepler problem
\begin{equation}
\ddot{\bmf r} + \frac{GM}{r^2} \bmf n = \bmf f,
\end{equation}
where $\bmf r = \bmf z_1 - \bmf z_2$ is the separation of the binary of total mass~$M$, $\bmf n = \bmf r/r$, and $\bmf f$ is an additional force (per unit mass) acting on the system, which we will treat as a small perturbation. After time-averaging over one orbit, the force $\bmf f$ results in a secular decay of the orbital period given to first order by
\begin{align}
\avg{\dot P}
=
\frac{3 GM}{(GM\Omega)^{4/3}} \int_0^{2\pi}\dx u
[(\bmf f \cdot \bm\lambda) \sqrt{1-e^2}  + (\bmf f \cdot \bmf n)\, e \sin u ],
\label{eq:bin_constraints_period_decay}
\end{align}
where $\Omega$ is the orbital frequency of the unperturbed binary and $e$ is its eccentricity. The unit vector $\bm\lambda$ points along the direction orthogonal to both $\bmf n$ and the binary's angular momentum vector. The trajectory along the orbit is parametrized by the eccentric anomaly $u$, which can be related to the coordinate time $t$ via Kepler's equation, $\Omega(t-t_0) = u- e\sin u$. The orbital parameter $t_0$ is called the time of pericenter passage and can be set to zero in this calculation without loss of generality.

The power counting rules established in Sec.~\ref{sec:bin_dynamics} can be used to infer that, of the five effects we calculated, the forces due to the halo's gravitational potential and accretion will provide the largest contributions to $\dot P$, since they scale with the most negative powers of $v$. For this reason, we concentrate only on these two effects. Respectively, they exert the forces
\begin{align}
\bmf f_\text{bkg}^i &= -R^i{}_{0j0}(t,\bmf 0)\, \bmf r^j,
\\
\bmf f_\text{acc} &= - \left(\frac{\delta\dot M_1}{M_1^2} + \frac{\delta\dot M_2}{M_2^2} \right) M \nu \dot{\bmf r} - \frac{G(\delta M_1 + \delta M_2)}{r^2} \bmf n.
\end{align}

\begin{table}
\caption{Parameters of the supermassive black hole binary in quasar OJ287, reproduced from Ref.~\cite{Dey:2018mjg}. Errors have been omitted for any quantity accurate to at least three significant figures. The intrinsic period $P$ is determined by rescaling the value measured on Earth by the scale factor $(1+z)^{-1}$ \cite{Sillanpaa:1988zz}. The uncertainty on $\dot P$ is at the $1\sigma$ level.}
\label{table:parameters_OJ287}
\begin{ruledtabular}
\begin{tabular}{lcl}
Parameter & & Value \\
\hline
Redshift & $z$ & 0.306 \\
Primary black hole mass & $M_1$ & $1.83 \times 10^{10}~M_\odot$ \\
Secondary black hole mass & $M_2$ & $1.50 \times 10^8~M_\odot$ \\
Primary dimensionless spin parameter & $\chi_1$ & 0.381 \\
Eccentricity & $e$ & 0.657 \\
Intrinsic orbital period
& $P$ & $9.24~\text{yr}$ \\
Orbital period decay & $\dot P$ & $(99 \pm 6) \times 10^{-5}$
\end{tabular}
\end{ruledtabular}
\end{table}

Substituting these forces into Eq.~\eqref{eq:bin_constraints_period_decay}, we obtain an expression for $\avg{\dot P}$ that is a function of the local density $\rho$ we wish to constrain, the known orbital parameters as summarized in Table~\ref{table:parameters_OJ287}, and one unknown: the phase factor $\Upsilon$ of the background relative to our zero of our time. Not knowing what value this parameter ought to have, we can obtain a conservative estimate for $\avg{\dot P}$ by marginalizing over $\Upsilon$ assuming a uniform prior.\footnote{By randomly sampling values of $\Upsilon \in [0,2\pi)$ and observing how they affect the value of $\avg{\dot P}$, we have verified that our assumption of a uniform prior does not bias our conclusions.} The resulting expectation value is
\begin{equation}
\mathbb E[\avg{\dot P}] = \frac{1}{2\pi}\int_0^{2\pi}\dx\Upsilon \avg{\dot P}.
\end{equation}
This procedure automatically excludes any contribution from $\bmf f_\text{bkg}$, since the Riemann tensor is proportional to ${\cos (2\mu t + 2\Upsilon)}$. It is still possible to extract a meaningful constraint by choosing $\Upsilon$ such that we calculate the maximum possible value of $|\avg{\dot P}|$ (as Ref.~\cite{Blas:2016ddr} does for binary pulsars), but we will not elect to do so and will instead simply concentrate on $\bmf f_\text{acc}$. It turns out that the constraint we derive from $\bmf f_\text{acc}$ is several orders of magnitude better than what we would get from $\bmf f_\text{bkg}$. This is because $\bmf f_\text{acc}$ has a component (${\propto-\dot{\bmf r}}$) that is always opposing the binary's motion.

The contribution from $\bmf f_\text{acc}$ to the orbital period decay is
\begin{equation}
\mathbb E[\avg{\dot P}_\text{acc}] = -\frac{48 \pi G^2 M P (2\nu - e) \rho}{1-e}.
\label{eq:bin_exp_dot_P}
\end{equation}
Requiring that this have a magnitude less than $|\delta\dot P| = 6 \times 10^{-5}$ imposes the upper bound
\begin{equation}
\label{eq:bin_constraints_rho}
\rho \lesssim 2 \times 10^9~M_\odot \,\text{pc}^{-3}
\end{equation}
at the $1\sigma$ level for the local density of FDM. Note that Eq.~\eqref{eq:bin_exp_dot_P} assumes that the black holes are spherical for simplicity (even though the spin of the primary black hole has been measured), since that is good enough for deriving an order-of-magnitude constraint. As a final step, it is necessary to check that this bound is consistent with the IR cutoffs in Eqs.~\eqref{eq:bin_smbh_instantaneous_scalar_condition} and \eqref{eq:bin_smbh_est_max_a}. While the second is easily satisfied for the case of OJ287, which has an orbital separation ${a \approx 56~\text{mpc}}$, the first of these tells us that our conclusions are valid only for scalars with a mass $\mu \ll 8 \times 10^{-22}~\text{eV}$.

The constraint in Eq.~\eqref{eq:bin_constraints_rho} is very weak, as FDM halos are expected to have core densities of around $100~M_\odot\,\text{pc}^{-3}$ \cite{Schive:2014dra,Schive:2014hza,Bar:2018acw,Boskovic:2018rub}. Accordingly, we conclude that typical dark matter halos are too dilute to leave any observable imprints in the inspiral of a black hole binary. This is entirely in line with our expectations going in. Nonetheless, the work in this section is still useful for illustrating how our EFT framework can be used to make quantitative predictions. In the next section, we will briefly comment on other scalar-field environments with greater observational potential that are worth exploring in future work.


\section{Conclusions}
\label{sec:conclusion}

We have developed a worldline EFT that accurately describes how black holes in general relativity interact with minimally coupled, real scalar fields. Stringent no-hair theorems limit the kinds of terms that are allowed in the effective action---in particular, black holes are not permitted any permanent scalar multipole moments of their own---but we still uncover a rich phenomenology when accounting for finite-size effects. Being an extension of Goldberger and Rothstein's construction \cite{Goldberger:2004jt,Goldberger:2005cd}, the novelty of our approach is in the integrating out of composite operators localized on the worldline, which encode information about UV physics transpiring near the horizon. This procedure proved to be a powerful method for generating new terms in the effective action never before considered in the literature. Central to this achievement was our use of the in-in formalism of quantum field theory, which enabled the accounting of dissipative effects at the level of the action.

Our EFT reveals that the motion of a black hole embedded in a scalar-field environment exhibits three features that distinguish it from other compact objects: First, the black hole experiences a drag force due to accretion of the background scalar field, which proceeds at a rate that is uniquely determined (at leading order) by the area of its horizon. Second, a scalar-field environment induces a scalar charge onto the black hole, granting it the ability to radiate energy and momentum into scalar waves. Third, the onset of this scalar charge also stipulates that a black hole must move under the influence of a fifth force.

Of these three effects, accretion is the most natural and unsurprising. Accordingly, many studies \cite{Bar:2018acw,Ferreira:2017pth,Macedo:2013qea} have appreciated its importance, which in optimal scenarios may even dominate over radiation reaction in driving the evolution of a black hole's inspiral \cite{Macedo:2013qea}. However, typical estimates for the accretion rate often rely on the absorption cross section for free, collisionless, nonrelativistic particles \cite{Unruh:1976fm}, which is strictly valid only for a black hole moving slowly through a gas of such particles. In contrast, we are often more interested in the motion of a black hole through a background field that is localized and bound by its own self-gravity. To qualify as a background, the total mass of this configuration must also be much greater than that of the black hole. In such cases, the correct accretion rate is determined from computing the flux of this scalar field across the horizon \cite{Jacobson:1999vr,Gregory:2017sor,Gregory:2018ghc,UrenaLopez:2002du}. As we pointed out earlier, what is remarkable is that this accretion rate emerges naturally from first principles in our EFT. Importantly, our equation for the resulting drag force works not only in the Newtonian regime, but holds in a fully relativistic setting.

Less obvious is the fact that black holes gain scalar charges when embedded in a scalar-field environment. The prediction of scalar radiation originates with Horbatsch and Burgess \cite{Horbatsch:2011ye}, but to the best of our knowledge, we are the first to point out that a black hole can experience a fifth force mediated by a \emph{minimally coupled} scalar field. While scalar radiation and fifth forces are par for the course in alternative theories of gravity, owing to a nonminimal coupling between the scalar and one or more curvature tensors \cite{Fujii:2003pa,Clifton:2011jh,Will:2014kxa}, the effects discussed in this paper emerge as necessary and inescapable consequences of accretion of the background scalar onto the black hole. Our EFT exposes this connection in no uncertain terms, showing that all three effects---accretion, scalar radiation, and the fifth force---can be traced back to a single parent term in the effective action.

We illustrated how this EFT can be used to make quantitative predictions by studying the early inspiral of a black hole binary located in the core of a fuzzy dark matter halo. This example was useful as a case study, since a series of approximations made performing calculations straightforward, but ultimately, typical halos in these models are too dilute to leave any observable imprints in the binary's inspiral. This is no cause for discouragement, however, as there are still other examples of scalar-field environments worth studying, which may have greater observational potential. At least two come to mind: Even if an ultralight scalar field is not produced in large abundances during the early Universe, rapidly rotating black holes with radii coincident with the scalar's Compton wavelength can quickly generate a corotating condensate of the field through a superradiant instability \cite{Brito:2015oca,Arvanitaki:2010sy,Arvanitaki:2014wva,Arvanitaki:2016qwi,Yoshino:2013ofa,*Yoshino:2014wwa,Brito:2014wla}. Such a system is outside the regime of validity of our point-particle EFT, since there is no separation of scales between the scalar condensate and its host black hole, but our EFT is perfectly poised to study what would happen to a much smaller black hole orbiting this system. In more exotic scenarios, it is also possible to envision a stellar-mass black hole orbiting a supermassive, compact horizonless object like a boson star \cite{Liebling:2012fv}. Both of these extreme-mass-ratio inspiral scenarios have been studied in the past \cite{Macedo:2013qea,Ferreira:2017pth}, albeit using a Newtonian approach with finite-size effects included in an \emph{ad hoc} fashion. Our EFT provides a systematic framework for extending these results into the fully relativistic regime while also accounting for effects associated with the black hole's induced scalar charge, hitherto unexplored. This points to one exciting direction for future work.

Also in the future, it will be interesting to extend our EFT to include a black hole's spin and to push its capabilities beyond leading, nontrivial order in the separation-of-scale parameters. The novel techniques we have employed when constructing the effective action are also likely to be invaluable when modeling the interactions of black holes or other compact objects with external scalar, vector, or tensor fields.

\begin{acknowledgments}
It is a pleasure to thank Cliff Burgess, Vitor Cardoso, Bogdan Ganchev, Joe Keir, Jorge Santos, Ulrich Sperhake, and Ira Rothstein for helpful comments and discussions.
This work has been partially supported by STFC Consolidated Grants No.~ST/P000371/1, No.~ST/P000673/1, and No.~ST/P000681/1.
L.K.W. is supported by the Cambridge Commonwealth, European and International Trust, and Trinity College, Cambridge.
R.G. is also supported in part by Perimeter Institute. Research at Perimeter Institute is supported by the Government of Canada through the Department of Innovation, Science and Economic Development and by the Province of Ontario through the Ministry of Research and Innovation.
\end{acknowledgments}

\appendix
\section{Deriving\protect\\the point-particle action}
\label{sec:rules_worldline}

This Appendix collates several technical details used in deriving the point-particle action $S_p$ in Sec.~\ref{sec:eft}.

\subsection{Two-point correlation functions}
Let us review several key features of two-point correlation functions. The basic ingredients are the Wightman functions
\begin{subequations}
\label{eq:eft_chi_definitions}
\begin{align}
-i\chi_+^{LL'}(\tau,\tau') &\coloneq \avg{ q^L(\tau) q^{L'}(\tau')},
\\
-i\chi_-^{LL'}(\tau,\tau') &\coloneq \avg{q^{L'}(\tau')q^L(\tau)},
\end{align}
from which all other two-point functions can be built. Respectively, we define the Feynman, Dyson, Hadamard, and Pauli-Jordan propagators as
\begin{align}
-i\chi_F^{LL'}(\tau,\tau') &\coloneq \avg{T q^L(\tau)q^{L'}(\tau')},
\allowdisplaybreaks\\
-i\chi_D^{LL'}(\tau,\tau') &\coloneq \avg{T^* q^L(\tau) q^{L'}(\tau')},
\allowdisplaybreaks\\
-i\chi_H^{LL'}(\tau,\tau') &\coloneq \avg{\{q^L(\tau),q^{L'}(\tau')\}},
\allowdisplaybreaks\\
-i\chi_C^{LL'}(\tau,\tau') &\coloneq \avg{[q^L(\tau),q^{L'}(\tau')]},
\end{align}
where $T$ and $T^*$ denote the time-ordering and anti-time-ordering operators, respectively. Whether minus signs or factors of $i$ appear on the lhs is simply a matter of convention. Note also that $-i\chi_C$ is nothing but the commutator. Last but not least, we define the retarded and advanced propagators by
\begin{align}
\label{eq:def_chi_R}
\chi_R^{LL'}(\tau,\tau') &\coloneq \theta(\tau-\tau') \chi_C^{LL'}(\tau,\tau'),
\\
\chi_A^{LL'}(\tau,\tau') &\coloneq -\theta(\tau'-\tau) \chi_C^{LL'}(\tau,\tau'),
\end{align}
\end{subequations}
where $\theta(x)$ is the Heaviside step function.

Not all of these two-point functions are independent. Notice from their definitions that
\begin{subequations}
\label{eq:eft_chi_identities}
\begin{equation}
\chi_+^{LL'}(\tau,\tau') = \chi_-^{L'L}(\tau',\tau)
\end{equation}
and, furthermore, the identity $\theta(x)+\theta(-x)=1$ implies
\begin{align}
\chi_R &= \chi_F - \chi_- = \chi_+ - \chi_D,
\\
\chi_A &= \chi_F - \chi_+ = \chi_- - \chi_D,
\\
\chi_H &= \chi_F + \chi_D = \chi_+ + \chi_-.
\end{align}
In these last three equations, all two-point functions have the same indices $LL'$ and arguments~$(\tau,\tau')$, which have been suppressed for readability. 
\end{subequations}

\subsection{Charge density}

In the main text, we defined the induced charge density of the black hole as
\begin{equation}
\Q^A(x) \coloneq \frac{1}{\sqrt{-g}}\int_{x'} \X^{A+}(x,x')\Phi(x').
\end{equation}
To obtain the end result in Eq.~\eqref{eq:eft_worldline_vertices_scalar_PL}, we substitute in explicit expressions for $\X^{A+}$ and simplify. The two cases $A \in \{+,-\}$ must be treated separately, but since the steps are almost identical, it suffices to work through just one example. Let us do~$\Q^+$. Using Eq.~\eqref{eq:eft_X_explicit}, we obtain
\eqskip
\begin{align}
\sqrt{-g}\Q^+ = &\int_{\lambda,\lambda'} \{
\left[ \chi_R(\lambda,\tau_{1'}) \Delta_1 - \chi_C(\lambda,\tau_{1'}) \Delta_+ \right]\dot\tau_{1'} \Phi(z_{1'})
\nonumber\\
&-
\left[ \chi_R(\lambda,\tau_{2'}) \Delta_2 - \chi_C(\lambda,\tau_{2'}) \Delta_+ \right]\dot\tau_{2'} \Phi(z_{2'})
\}
\end{align}
after integrating over the delta functions in~$\Delta_{a'}$. We write $\Delta_+ = (\Delta_1+\Delta_2)/2$, $\tau_{a'}\equiv \tau_a(\lambda')$, and $z_{a'}\equiv z_a(\lambda)$ for brevity. Now substitute in explicit forms for $\chi_R$ and $\chi_C$, given by Eq.~\eqref{eq:eft_worldline_vertices_chi_R}, to obtain
\begin{gather}
\sqrt{-g}\Q^+ = \A \int_{\lambda,\lambda'}\int_\w
[\Delta_1 \dot\tau_{2'}i\w e^{-i\w(\lambda-\tau_{2'})}\Phi(z_{2'}) - (1\leftrightarrow 2)].
\end{gather}
Recognizing that $\dot\tau_{2'} i\w$ can be rewritten as a derivative $\dx/\dx\lambda'$ acting on the exponential, and likewise for $1\leftrightarrow 2$, we find
\begin{equation}
\sqrt{-g}\Q^+ =
-\A \int_{\lambda,\lambda'}
[\Delta_1 \dot\Phi(z_{2'}) \delta(\lambda-\tau_{2'}) - (1\leftrightarrow 2)]
\end{equation}
after integrating by parts. Finally, integrating over the remaining delta functions in $\Delta_{1,2}$ gives us
\begin{gather}
\Q^+ = - \A\int_\lambda \frac{\delta^{(4)}(x-z_1)}{\sqrt{-g}} \dot\tau_1 \int_{\lambda'}\dot\Phi(z_{2'}) \delta(\tau_1 - \tau_{2'}) - (1\leftrightarrow 2).
\end{gather}

Repeating similar steps to obtain an expression for $\Q^-$, we recognize the following pattern: If we define
\begin{align}
\Q_1(x) \coloneq &
- \A \int_\lambda \frac{\delta^{(4)}\bm( x-z_1(\lambda) \bm)}{\sqrt{-g}} \dot\tau_1(\lambda)
\int_{\lambda'} \dot\Phi\bm( z_2(\lambda') \bm)
\nonumber\\
&\times \delta\bm( \tau_1(\lambda)-\tau_2(\lambda') \bm)
\end{align}
and define $\Q_2(x)$ by interchanging ${1\leftrightarrow 2}$ in the above equation, then the charge densities $\Q^{\mp} \equiv \Q_\pm$ in the Keldysh representation are obtained through the usual transformation rule in Eq.~\eqref{eq:def_Keldysh_repn}.

\subsection{Accretion rate}
We now turn to deriving the accretion rate. Our starting point is
\begin{equation}
S_p \supset \frac{1}{2} \int_{\lambda,\lambda'} \dot\tau_1\dot\tau_{2'} \chi_C(\tau_1,\tau_{2'})\Phi(z_1)\Phi(z_{2'}).
\end{equation}
As we did in the main text, we perturb the proper time such that $\tau_a \to \tau_a + \delta\tau_a$. The terms linear in $\delta\tau_a$ are
\begin{align}
\frac{1}{2}&\int_{\lambda,\lambda'} \dot\tau_1\dot\tau_{2'} \chi_C(\tau_1,\tau_{2'})\Phi(z_1)\Phi(z_{2'})
\nonumber\\&\times
\left[
\left( \frac{\delta\dot\tau_1}{\dot\tau_1} + \frac{\partial_{(1)}\chi_C}{\chi_C} \int_{\lambda_i}^\lambda \dx\sigma \delta\dot\tau_1(\sigma) \right)
\right.\nonumber\\
&+ \left.
\left( \frac{\delta\dot\tau_{2'}}{\dot\tau_{2'}} + \frac{\partial_{(2)}\chi_C}{\chi_C} \int_{\lambda_i}^{\lambda'} \dx\sigma \delta\dot\tau_2(\sigma) \right)
\right],
\end{align}
where we write $\partial_{(n)}$ to mean the derivative with respect to the $n$th argument. Using the explicit expression for $\chi_C$ in Fourier space, this becomes
\eqskip
\begin{align}
\A &\int_{\lambda,\lambda'} \Phi(z_1)\Phi(z_{2'})
\int_\w e^{-i\w(\tau_1 - \tau_{2'})}
\nonumber\\&\times
\left[
\left( \delta\dot\tau_1 \dot\tau_{2'}i\w - \dot\tau_1 \dot\tau_{2'}(i\w)^2 \int_{\lambda_i}^\lambda \dx\sigma \delta\dot\tau_1(\sigma) \right)
\right.
\nonumber\\&
+
\left.
\left( \delta\dot\tau_{2'} \dot\tau_1 i\w + \dot\tau_1 \dot\tau_{2'}(i\w)^2 \int_{\lambda_i}^{\lambda'} \dx\sigma \delta\dot\tau_2(\sigma) \right)
\right].
\label{eq:rules_worldline_accretion_rate_intermediate}
\end{align}
Just as we did when deriving the charge density, recognize that each appearance of $\dot\tau_1 i\w$ can be replaced by a derivative $-\dx/\dx\lambda$ acting on the exponential, and likewise each factor of $\dot\tau_{2'}i\w$ can be replaced by $\dx/\dx\lambda'$. Having done so, Eq.~\eqref{eq:rules_worldline_accretion_rate_intermediate} simplifies to
\begin{align}
\A &\int_{\lambda,\lambda'} \Phi(z_1)\Phi(z_{2'})
\left[
\frac{\dx}{\dx\lambda}\left( \int_{\lambda_i}^\lambda \dx\sigma \delta\dot\tau_1(\sigma) \frac{\dx}{\dx\lambda'} \right)
\right.
\nonumber\\
&\left.
-\frac{\dx}{\dx\lambda'}\left( \int_{\lambda_i}^{\lambda'}\dx\sigma \delta\dot\tau_2(\sigma) \frac{\dx}{\dx\lambda} \right)
\right]
\delta(\tau_1 - \tau_{2'}).
\end{align}
Integrating by parts then yields
\begin{align}
\A &\int_{\lambda_i}^{\lambda_f}\dx\lambda \int_{\lambda_i}^{\lambda}\dx\sigma \delta\dot\tau_1(\sigma) \dot\Phi(z_1)
\int_{\lambda'} \dot\Phi(z_{2'}) \delta(\tau_1 - \tau_{2'})
\nonumber\\
&- (1\leftrightarrow 2).
\end{align}
Note that $(\lambda_i,\lambda_f)$ correspond to the initial and final times at which boundary conditions are to be specified according to the in-in formalism. The final result in Eq.~\eqref{eq:eft_worldline_vertices_accretion_formula} is obtained after swapping the integration limits on $\lambda$ and $\sigma$ by using the identity
\begin{align}
\int_{\lambda_i}^{\lambda_f} \dx\lambda \int_{\lambda_i}^\lambda\dx\sigma
&=
\int_{\lambda_i}^{\lambda_f} \dx\sigma \int_\sigma^{\lambda_f}\dx\lambda
\nonumber\\
&=
\int_{\lambda_i}^{\lambda_f} \dx\sigma \left( \int_{\lambda_i}^{\lambda_f} \dx\lambda - \int_{\lambda_i}^\sigma \dx\lambda \right).
\end{align}

\section{Propagators and bulk\protect\\vertices in weakly curved spacetimes}
\label{sec:rules_bulk}

As we did for the point-particle action in Sec.~\ref{sec:eft_worldline_vertices}, we substitute the decomposition~\eqref{eq:eft_matching_field_decomposition} into the field action $S_f$ to obtain the series
\begin{equation}
S_f = \sum_{n_h=0}^\infty \sum_{n_\varphi=0}^\infty S_f^{(n_h + n_\varphi)},
\end{equation}
where recall the integers $(n_h,n_\varphi)$ count the powers of the field perturbations appearing in each term. Since the background $(g,\Phi)$ is assumed to be a valid solution of the field equations, there are no terms with $n_h + n_\varphi < 2$. With general relativity being a gauge theory, it is necessary that we supplement $S_f$ with a gauge-fixing term \emph{\`{a} la} Faddeev and Popov,
\begin{equation}
S_\text{gf} = -\int_x \sqrt{-g} c_{AB} g^{\mu\nu} G_\mu^A G_\nu^B,
\end{equation}
which imposes the gauge condition $G_\mu^A \approx 0$. If we impose the generalized Lorenz gauge 
\begin{equation}
\label{eq:rules_bulk_Lorenz_gauge}
G_\mu^A = \nabla^\nu\left( h^A_{\mu\nu} - \frac{1}{2} h^A g_{\mu\nu}\right) - \frac{\zeta}{2\mpl}\varphi^A \nabla_\mu\Phi
\end{equation}
defined in terms of an arbitrary constant $\zeta$, the part of the field action quadratic in the perturbations is
\begin{align}
S_f^{(2)} =
&\;\frac{1}{2}c_{AB}\int_x\sqrt{-g}\bigg\{
h_{\alpha\beta}^A (P^{\alpha\beta\mu\nu}\Box - \mathcal M^{\alpha\beta\mu\nu}) h_{\mu\nu}^B
\nonumber\\
&+ \frac{1-\zeta}{2\mpl} h^A_{\mu\nu}
[2 \nabla^\mu\Phi \nabla^\nu - g^{\mu\nu}(\nabla_\alpha\Phi \nabla^\alpha + \mu^2\Phi)]
\varphi^B
\nonumber\\
&- \frac{\zeta}{\mpl} h^A_{\mu\nu}\varphi^B \nabla^\mu\nabla^\nu\Phi + \varphi^A(\Box-\mu^2_\text{eff})\varphi^B
\bigg\}.
\label{eq:rules_bulk_Sf_quadratic}
\end{align}
This is expressed in terms of three background quantities:
\begin{subequations}
\begin{align}
P^{\alpha\beta\mu\nu} &=
\frac{1}{2} \big( g^{\alpha\mu}g^{\beta\nu} + g^{\alpha\nu}g^{\beta\mu} - g^{\alpha\beta} g^{\mu\nu} \big),
\allowdisplaybreaks\\
\mathcal M^{\alpha\beta\mu\nu} &=
2 \big(g^{\alpha\mu} R^{\beta\nu} - R^{\alpha\mu\beta\nu} \big)
- \frac{\mu^2\Phi^2}{4\mpl^2} P^{\alpha\beta\mu\nu},
\allowdisplaybreaks\\
\mu^2_\text{eff} &= \mu^2 + \frac{2\zeta^2}{\mpl^2}\nabla_\alpha\Phi \nabla^\alpha\Phi.
\end{align}
\end{subequations}

The convenient gauge choice $\zeta = 1$ exchanges derivative interactions between the different field perturbations in favor of simpler algebraic ones, but nonetheless an arbitrary background with $\Phi\neq 0$ will lead to a quadratic action that mixes the graviton with the scalar. In general, these mixing terms must be treated nonperturbatively, meaning we cannot speak of a propagator for $h$ and a separate propagator for~$\varphi$ \cite{Zimmerman:2015rga}.

An exception to this rule is when $\Phi\sim\mathcal O(\eps)$ is itself a weak perturbation living on top of a vacuum geometry. In such cases, the background solution admits its own expansion in the small bookkeeping parameter $\eps$, namely
\begin{equation*}
\Phi = \Phi^{(1)} + \mathcal O(\eps^2),\quad
g = g^{(0)} + g^{(2)} + \mathcal O(\eps^3).
\end{equation*}
The example of a fuzzy dark matter halo we consider in Sec.~\ref{sec:bin} admits this expansion; the vacuum spacetime is described by the Minkowski metric, ${g^{(0)} = \eta}$, which is only weakly perturbed by the gravitational potential ${g^{(2)} \equiv H}$ of the halo. This Appendix establishes the Feynman rules for backgrounds of this form.

\subsection{Free-field propagators}

Since ${\Phi \sim \mathcal O(\eps)}$ is assumed to be small, the mixing terms in the second and third lines of Eq.~\eqref{eq:rules_bulk_Sf_quadratic} can be treated perturbatively as interactions. Hence, the graviton and scalar now have their own propagators, which are defined on flat space. The gauge-fixed generating functional for the free fields, which we introduced in Sec.~\ref{sec:eft_matching}, is then
\begin{align}
\mathcal Z_0[j,J] = &\exp \left( \frac{i}{2} \int_{x,x'} J^A_{\alpha\beta}(x) D_{AB}^{\alpha\beta\mu\nu}(x,x') J^B_{\mu\nu}(x') \right)
\nonumber\\
&\times \exp\left( \frac{i}{2} \int_{x,x'} j^A(x) G_{AB}(x,x') j^B(x') \right).
\end{align}
Directly analogous to Eq.~\eqref{eq:def_chi_AB}, the scalar field has a matrix of propagators given by
\begin{equation}
G_{AB} =
\begin{pmatrix}
\frac{1}{2} G_H & G_R \\
G_A & 0
\end{pmatrix}
\end{equation}
in the Keldysh representation, whereas the matrix of graviton propagators reads
\begin{equation}
D_{AB}^{\alpha\beta\mu\nu} = P^{\alpha\beta\mu\nu}
\begin{pmatrix}
\frac{1}{2} D_H & D_R \\
D_A & 0
\end{pmatrix}
\end{equation}
in the Lorenz gauge $\zeta = 0$.\footnote{We must set $\zeta = 0$ when ${\Phi \sim \mathcal O(\eps) \ll 1}$, as the last term in Eq.~\eqref{eq:rules_bulk_Lorenz_gauge} is now smaller than the others. Were we to choose a nonzero value for $\zeta$, the gauge-fixing term would still attempt to enforce the Lorenz gauge at $\mathcal O(\eps^0)$ and then subsequently attempt to enforce the condition $\varphi^A \nabla_\mu\Phi \approx 0$ at $\mathcal O(\eps^1)$. This would be undesirable.} Note that the tensor $P$ is here defined in terms of the Minkowski metric.

One finds that both propagator matrices are symmetric under the simultaneous interchange of the arguments $x \leftrightarrow x'$ and the CTP indices $A \leftrightarrow B$. As a result, appropriate relabeling of dummy indices and integration variables can always be used to replace the advanced propagators $(G_A,D_A)$ in a Feynman diagram with the retarded propagators $(G_R,D_R)$. Moreover, much like with the Hadamard propagator $\chi_H$ for the worldline operators in Sec.~\ref{sec:eft_integration_dynamical_worldlines}, one also finds that $(G_H,D_H)$ are always flanked by at least two quantities that vanish in the physical limit. Consequently, they never contribute to the (classical) equations of motion \cite{Galley:2008ih,Galley:2009px}, and can therefore be neglected. Taken together, these observations tell us that only the retarded propagators are needed for calculating physical observables. For our purposes, it is most convenient to write the scalar's retarded propagator as
\begin{equation}
G_R(x,x') = \int_p \frac{e^{i p \cdot (x-x')}}{-(p^0 + i\epsilon)^2 + \bmf p^2 + \mu^2}.
\end{equation} 
The graviton's retarded propagator $D_R$ has an identical expression, except with $\mu = 0$.

\subsection{Bulk vertices}

\begin{figure}
\centering\includegraphics{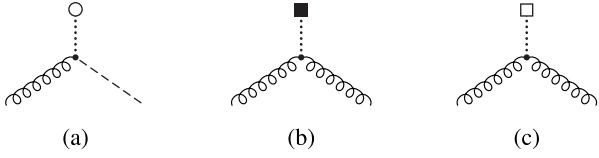}
\caption{Examples of bulk vertices. The graviton $h$ is drawn as a helical line while the scalar $\varphi$ is drawn as a dashed line. Insertions of the background fields are denoted as dotted lines terminating in a given shape. The circle, filled square, and empty square correspond to the background scalar $\Phi$, the mass tensor $\mathcal M$, and the background gravitational potential $H$, respectively.}
\label{fig:bulk_vertices}
\end{figure}

We treat every term in the field action $S_f$ not included in the generating functional $\mathcal Z_0$ perturbatively as an interaction vertex. Three are relevant for the purposes of this paper. At linear order in $\eps$, the aforementioned mixing terms give us the vertex
\begin{equation}
\mathcal L_{h\varphi\Phi} = \frac{h_A^{\mu\nu}}{2\mpl}
[2 \partial_\mu\Phi \partial_\nu - \eta_{\mu\nu}(\partial^\alpha\Phi \partial_\alpha + \mu^2\Phi)]\varphi_A,
\end{equation}
drawn in Fig.~\ref{fig:bulk_vertices}(a). The second vertex, depicted in Fig.~\ref{fig:bulk_vertices}(b), is an effective mass for the graviton,
\begin{equation}
\mathcal L_{h^2\mathcal M} = - \frac{1}{2} h^A_{\alpha\beta} \mathcal M^{\alpha\beta\mu\nu} h_{A,\mu\nu},
\end{equation}
where the mass tensor ${\mathcal M \sim \mathcal O(\eps^2)}$ at leading order. The final vertex in Fig.~\ref{fig:bulk_vertices}(c) comes from expanding the background metric in the graviton's kinetic term to first order in $H$:
\begin{align}
\mathcal L_{h^2H} = \frac{1}{2} H_{\mu\nu}
\left[
\frac{1}{\sqrt{-g}} \frac{\partial}{\partial g_{\mu\nu}} \left(\sqrt{-g} h^A_{\alpha\beta}(P^{\alpha\beta\rho\sigma}\Box)h_{A,\rho\sigma} \right)
\right]_{g = \eta}.
\nonumber\\
\end{align}
This generates a large number of terms that derivatively couple $H_{\mu\nu}$ to the gravitons. We will omit writing down an explicit expression, since it will not be needed for any of our calculations in this paper.

\subsection{Position-space Feynman rules}

Let us schematically denote each bulk vertex as
\begin{equation}
S_f \supset \int_x V_f \frac{h^{n_h}}{n_h!}  \frac{\varphi^{n_\varphi}}{n_\varphi!},
\end{equation}
where all indices have been suppressed. In general, $V_f$ is a derivative operator acting on the fields. The worldline vertices in Sec.~\ref{sec:eft_worldline_vertices} are denoted in a similar way by replacing subscript $f$'s with subscript $p$'s. The position-space Feynman rules for our EFT are then as follows:
\begin{enumerate}
\item Each bulk vertex gives an appropriate factor of $i V_f$, while each worldline vertex gives a factor of $i V_p$.
\item Each graviton or scalar line corresponds to an appropriate propagator matrix, either $-i D_{AB}^{\alpha\beta\mu\nu}$ or $-i G_{AB}$, respectively.
\item All CTP and spacetime indices are to be summed over, and all spacetime points are to be integrated over, except those corresponding to external legs.
\item Divide each diagram by the appropriate symmetry factor.
\end{enumerate}
If the diagram being computed has no external legs, we choose to additionally multiply by a factor of $-i$ such that it constitutes a term in the effective action $\Gamma$ rather than one in $i\Gamma$.

\section{Master integral}
\label{sec:master_integral}

Many occasions in the main text call for the evaluation of an integral of the form
\begin{equation}
\mathcal G[f](t,\bmf x) \coloneq \int_{t'} G_R(t,\bmf x; t',\bmf 0) f(t'),
\end{equation}
which describes the leading-order expectation value for $\varphi$ due to a time-dependent source $f(t)$ at rest at the origin. In the interest of efficiency, let us discuss how to evaluate this integral (on flat space) once for an arbitrary source $f(t)$.

We begin by expressing both $G_R$ and $f$ in Fourier space to find
\begin{equation}
\mathcal G[f] =
\int_{t'} \int_\w \tilde f(\w) e^{-i\w t'} \int_p \frac{e^{-i p^0 (t-t')} e^{i \bmf p \cdot \bmf x}}{-(p^0 + i\epsilon)^2 + \bmf p^2 + \mu^2}.
\end{equation}
Integrating over $t'$ generates a delta function which imposes the condition $p^0 = \w$. Also integrating over $p^0$ then gives
\begin{equation}
\mathcal G[f] = \int_\w \tilde f(\w)e^{-i\w t} \int_{\bmf p} \frac{e^{i \bmf p \cdot \bmf x}}{\bmf p^2 + \mu^2 - (\w + i\epsilon)^2}.
\label{eq:rules_bulk_integral_p_branch_off}
\end{equation}

The integral over momentum space must be evaluated separately depending on the sign of the real part of $k^2 = (\w+i\epsilon)^2 - \mu^2$. When $k^2 \leq 0$, the scalar gives rise to the Yukawa potential
\begin{equation}
I(\w) \coloneq \int_{\bmf p} \frac{e^{i \bmf p \cdot \bmf x}}{\bmf p^2 - k^2}
\supset
\theta(-k^2) \frac{e^{- \sqrt{\mu^2-\w^2}r}}{4\pi r}.
\label{eq:rules_bulk_integral_w_negative}
\end{equation}
If instead $k^2 > 0$, we expect this equation to describe spherical waves emanating from the origin. Indeed, performing the integral yields
\begin{equation}
I(\w) \supset \frac{\theta(k^2)}{4\pi r}
\big( \theta(\w) e^{i\sqrt{\w^2-\mu^2}r} + \theta(-\w)e^{-i\sqrt{\w^2-\mu^2}r}
\big).
\label{eq:rules_bulk_integral_w_positive}
\end{equation}
The complete result for $I(\w)$ is formed by taking the sum of these two equations. Substituting this back into Eq.~\eqref{eq:rules_bulk_integral_p_branch_off}, we find that we can write
\begin{equation}
\mathcal G[f] = \frac{1}{4\pi r} \int_\w \tilde f(\w) e^{-i \w t + i k(\w) r},
\end{equation}
where the root of $k^2$ is defined as
\begin{equation}
k(\w) \coloneq
\begin{cases}
i \sqrt{\mu^2 - \w^2} & \w^2 \leq \mu^2 \\
\text{sgn}(\w) \sqrt{\w^2 - \mu^2} & \w^2 > \mu^2.
\end{cases}
\end{equation}

Occasionally, it will be convenient to simplify this further by exploiting the fact that the Fourier transform of a real source $f(t)$ satisfies $\tilde f(-\w) = \tilde f^*(\w)$, while our definition for $k(\w)$ satisfies $k(-\w) = -k^*(\w)$. Thus, an equivalent expression is
\begin{equation}
\mathcal G[f] = \frac{1}{4\pi r} 2\Real \int_0^\infty\frac{\dx\w}{2\pi}
\tilde f(\w) e^{-i\w t + i k(\w) r}.
\end{equation}


\section{An EFT approach\protect\\to dynamical friction}
\label{sec:derivation_dynamical_friction}

Here we derive the drag force in Eq.~\eqref{eq:bin_smbh_dynamical_friction} due to dynamical friction. Taking its nonrelativistic limit, the graviton vertex in Eq.~\eqref{eq:eft_worldline_vertices_graviton} reduces to
\begin{equation}
S_{p,\K} \supset \frac{M_\K}{2\mpl}\int_x \delta^A_\K(x) h_{A,00}(x)
\end{equation}
at leading order in $v$. This can be used in conjunction with the Feynman rules in Appendix~\ref{sec:rules_bulk} to obtain
\begin{align}
\text{Fig.~\ref*{fig:bin_smbh_dynamical_friction}(a)}
=&\;
\left(\frac{M}{2\mpl}\right)^2 \int_{x,x',y,y'} \delta^+(x)\delta^-(x')
\nonumber\\
&\times\int_{p,p',q}
\frac{e^{ip\cdot(x-y)}}{\bmf p^2}
\frac{e^{iq\cdot(y-y')}}{\bmf q^2}
\frac{e^{ip'\cdot(y'-x')}}{\bmf p'^2}
\nonumber\\
&\times
P_{00\alpha\beta} V^{\alpha\beta}_{h\varphi\Phi}(y^0,q^0)
P_{00\mu\nu}V^{\mu\nu}_{h\varphi\Phi}(y'^0,-q^0),
\label{eq:dynamical_friction_raw}
\end{align}
having kept only the instantaneous part of the propagators; cf.~Eqs.~\eqref{eq:bin_smbh_propagators_instantaneous_limit_scalar} and \eqref{eq:bin_smbh_propagators_instantaneous_limit_graviton}. We have also suppressed the index $\K$ since this force acts independently on each member of the binary. The vertex functions in the third line read
\begin{equation}
P_{00\alpha\beta} V^{\alpha\beta}_{h\varphi\Phi}(y^0,q^0)
=
\frac{1}{2\mpl}\left[ -2iq^0 \dot\Phi(y^0) - \mu^2\Phi(y^0) \right].
\label{eq:dynamical_friction_vertex_function}
\end{equation}
The two terms in square brackets scale with different powers of $GM\mu$ and $v$, but it will be instructive to keep both of them around in this derivation. It turns out that the second term $\mu^2\Phi$ provides no contribution whatsoever to the force.

We first simplify Eq.~\eqref{eq:dynamical_friction_raw} by performing a number of trivial integrations. Integrating over $p^0$ and $p'^0$ produces delta functions that enforce the conditions $y^0 = x^0 \equiv t$ and $y'^0 = x'^0 \equiv t'$, respectively. Moreover, integrating over $\bmf y$ and $\bmf y'$ enforces the conservation of 3-momentum along the entire diagram, $\bmf p = \bmf p' = \bmf q$. The result is
\begin{gather}
\text{Fig.~\ref*{fig:bin_smbh_dynamical_friction}(a)}
=
\frac{M^2}{16\mpl^4} \int_{x,x'} \delta^+(x)\delta^-(x') \int_{q}  \frac{e^{i q \cdot(x-x')}}{\bmf q^6} W(q^0;t,t'),
\end{gather}
with
$W(q^0;t,t') = [-2iq^0\dot\Phi(t) - \mu^2\Phi(t)] [ 2iq^0 \dot\Phi(t') - \mu^2\Phi(t') ].$
We perform the integral over $\bmf q$ by utilizing the standard identity
\begin{equation}
\int \frac{\dx^d\bmf q}{(2\pi)^d} \frac{e^{i \bmf q \cdot \bmf r}}{(\bmf q^2)^\alpha}
=
\frac{1}{(4\pi)^{d/2}} \frac{\Gamma(d/2-\alpha)}{\Gamma(\alpha)}
\left( \frac{\bmf r^2}{4} \right)^{\alpha-d/2},
\end{equation}
while the integral over $q^0$ is performed by replacing each factor of $i q^0$ in $W(q^0;t,t')$ with a derivative $\dx/\dx t'$ acting on $e^{-i q^0(t-t')}$ and then integrating by parts. These steps give us
\begin{align}
\text{Fig.~\ref*{fig:bin_smbh_dynamical_friction}(a)}
&=
\frac{M^2}{16\mpl^4} \int_{t,t'} \delta(t-t') \left( W_0(t,t') + W_1(t,t') \frac{\dx}{\dx t'} \right.
\nonumber\\
&\quad\left. +\;  W_2(t,t') \frac{\dx^2}{\dx t'^2} \right)
\int_{\bmf x,\bmf x'} \delta^+(x)\delta^-(x') |\bmf x - \bmf x'|^3,
\end{align}
with $W_2(t,t') = -4 \dot\Phi(t) \dot\Phi(t')$. Determining expressions for $W_0$ and $W_1$ will not be necessary.

Now expand $\delta^\pm(x)$ in powers of $\bmf z_-$ according to Eq.~\eqref{eq:bin_smbh_delta_function_expand}. To linear order in $\bmf z_-$, the term involving $W_2$ yields
\begin{align}
\text{Fig.~\ref*{fig:bin_smbh_dynamical_friction}(a)}
\supset &\;
- \frac{8\pi}{3} (GM)^2 \int_{t,t'}\delta(t-t')
\dot\Phi(t)\dot\Phi(t') \bmf z_-^i(t)
\nonumber\\
&\times
\frac{\dx^2}{\dx t'^2} \left[ \frac{\partial}{\partial \bmf z_+^i(t)} |\bmf z_+(t) - \bmf z(t')|^3
\right]_\text{PL}.
\label{eq:dynamical_friction_raw_2}
\end{align}
Evaluating the derivatives, the second line becomes
\begin{align}
\frac{6 (\bmf r \cdot \bmf v) \bmf v_i}{|\bmf r|} + \frac{3 \bmf v^2 \bmf r_i}{|\bmf r|} - \frac{3 (\bmf r \cdot \bmf v)^2 \bmf r_i}{|\bmf r|^3}
- 3|\bmf r| \bmf a_i - \frac{3(\bmf r \cdot \bmf a) \bmf r_i}{|\bmf r|},
\label{eq:dynamical_friction_derivatives}
\end{align}
where we write $\bmf r \equiv \bmf z(t) - \bmf z(t')$ for brevity. Defining $s = t-t'$, we Taylor expand
$\bmf r = s \bmf v(t) + s^2 \bmf a(t)/2 + \mathcal O(s^3)$
and substitute it back into Eq.~\eqref{eq:dynamical_friction_raw_2} to obtain
\begin{equation}
\text{Fig.~\ref*{fig:bin_smbh_dynamical_friction}(a)}
\supset
- 16\pi (GM)^2 \int_t \dot\Phi{}^2 \bmf v^2 \hat{\bmf v}\cdot \bmf z_-,
\label{eq:dynamical_friction_final}
\end{equation}
after integrating over $s$. Notice that only the $\mathcal O(s^0)$ terms contribute to the force because of the delta function $\delta(s)$. The desired result can already be read off from Eq.~\eqref{eq:dynamical_friction_final}, meaning that the terms involving $W_0$ and $W_1$ do not contribute. This is easy to see, since
\begin{align*}
W_0(t,t')\left[ \frac{\partial}{\partial \bmf z_+^i(t)} |\bmf z_+(t) - \bmf z(t')|^3\right]_\text{PL} &\sim \mathcal O(s^2),
\\
W_1(t,t')\frac{\dx}{\dx t'} \left[ \frac{\partial}{\partial \bmf z_+^i(t)} |\bmf z_+(t) - \bmf z(t')|^3\right]_\text{PL} &\sim \mathcal O(s).
\end{align*}

\bibliography{bh}
\end{document}